\documentclass[journal,a4paper,oneside,onecolumn]{IEEEtran} 
\usepackage{amsfonts,amssymb}
\usepackage[utf8]{inputenc} 
\usepackage[T1]{fontenc}
\usepackage{url}
\usepackage{ifthen}
\usepackage{cite}
\usepackage[cmex10]{amsmath}
\usepackage{mathabx}
\newtheorem{thm}{Theorem}
\newtheorem{lem}{Lemma}
\newtheorem{df}{Definition}
\newtheorem{rem}{Remark}
\newtheorem{example}{Example}

\usepackage{curve2e}
\usepackage{color}

\newcommand{\fC}{\mathfrak{C}}
\newcommand{\NN}{\mathbb{N}}
\newcommand{\C}{\mathcal{C}}
\newcommand{\E}{\mathcal{E}}
\newcommand{\F}{\mathcal{F}}
\newcommand{\I}{\mathcal{I}}
\newcommand{\J}{\mathcal{J}}
\newcommand{\M}{\mathcal{M}}
\newcommand{\cS}{\mathcal{S}}
\newcommand{\T}{\mathcal{T}}
\newcommand{\U}{\mathcal{U}}
\newcommand{\V}{\mathcal{V}}
\newcommand{\W}{\mathcal{W}}
\newcommand{\X}{\mathcal{X}}
\newcommand{\Y}{\mathcal{Y}}
\newcommand{\Z}{\mathcal{Z}}
\newcommand{\G}{\mathcal{G}}
\newcommand{\cc}{\boldsymbol{c}}
\newcommand{\mm}{\boldsymbol{m}}
\newcommand{\bp}{\boldsymbol{p}}
\newcommand{\bs}{\boldsymbol{s}}
\newcommand{\uu}{\boldsymbol{u}}
\newcommand{\vv}{\boldsymbol{v}}
\newcommand{\ww}{\boldsymbol{w}}
\newcommand{\xx}{\boldsymbol{x}}
\newcommand{\yy}{\boldsymbol{y}}
\newcommand{\zz}{\boldsymbol{z}}
\newcommand{\CC}{\boldsymbol{C}}
\newcommand{\UU}{\boldsymbol{U}}
\newcommand{\VV}{\boldsymbol{V}}
\newcommand{\WW}{\boldsymbol{W}}
\newcommand{\XX}{\boldsymbol{X}}
\newcommand{\YY}{\boldsymbol{Y}}
\newcommand{\ZZ}{\boldsymbol{Z}}
\newcommand{\zero}{\boldsymbol{0}}
\newcommand{\one}{\boldsymbol{1}}
\newcommand{\aalpha}{\boldsymbol{\alpha}}
\newcommand{\bbeta}{\boldsymbol{\beta}}
\newcommand{\bcF}{\boldsymbol{\F}}

\newcommand{\hU}{\widehat{U}}
\newcommand{\hZ}{\widehat{Z}}
\newcommand{\hW}{\widehat{W}}
\newcommand{\hu}{\widehat{u}}
\newcommand{\hWW}{\widehat{\WW}}
\newcommand{\hww}{\widehat{\ww}}

\newcommand{\tW}{\widetilde{W}}

\newcommand{\chu}{\widecheck{u}}
\newcommand{\chU}{\widecheck{U}}
\newcommand{\chW}{\widecheck{W}}
\newcommand{\chww}{\widecheck{\ww}}

\newcommand{\limn}{\lim_{n\to\infty}}
\newcommand{\limsupn}{\limsup_{n\to\infty}}
\newcommand{\pliminfn}{\operatornamewithlimits{\mathrm{p-liminf}}_{n\to\infty}}
\newcommand{\plimsupn}{\operatornamewithlimits{\mathrm{p-limsup}}_{n\to\infty}}
\newcommand{\Prod}{\operatornamewithlimits{\text{\LARGE $\times$}}}

\newcommand{\oH}{\overline{H}}
\newcommand{\uH}{\underline{H}}
\newcommand{\uI}{\underline{I}}
\newcommand{\oI}{\overline{I}}
\newcommand{\oT}{\overline{\T}}
\newcommand{\uT}{\underline{\T}}
\newcommand{\od}{\overline{d}}
\newcommand{\oO}{\overline{O}}
\newcommand{\ugamma}{\underline{\gamma}}
\newcommand{\ogamma}{\overline{\gamma}}

\newcommand{\lrB}[1]{\left[{#1}\right]}
\newcommand{\lrb}[1]{\left\{{#1}\right\}}
\newcommand{\lrsb}[1]{\left({#1}\right)}
\newcommand{\lrbar}[1]{\left|{#1}\right|}

\newcommand{\markov}{\leftrightarrow}
\newcommand{\e}{\varepsilon}
\newcommand{\vphi}{\varphi}

\newcommand{\Prob}{\mathrm{Porb}}
\newcommand{\Error}{\mathrm{Error}}
\newcommand{\im}{\mathrm{Im}}

\newcommand{\ROP}{\mathcal{R}_{\mathrm{OP}}}
\newcommand{\RIT}{\mathcal{R}_{\mathrm{IT}}}
\newcommand{\RCRNG}{\mathcal{R}_{\mathrm{CRNG}}}
\newcommand{\helper}{\mathrm{helper}}

\newcommand{\Ipc}{\I'^{\complement}}

\allowdisplaybreaks

\title{
 Distributed Source Coding\\
 Using Constrained-Random-Number Generators
}
\author{
 \IEEEauthorblockN{Jun~Muramatsu~\IEEEmembership{Senior Member,~IEEE}}
 \\
  \IEEEauthorblockA{
   NTT Communication Science Laboratories, NTT Corporation\\
   2-4 Hikaridai, Seika-cho, Soraku-gun, Kyoto 619-0237, Japan\\
   E-mail: jun.muramatsu@ieee.org.
  }
}
\begin{document}
\maketitle
\begin{abstract}
This paper investigates the general distributed lossless/lossy source coding
formulated by Jana and Blahut.
Their multi-letter rate-distortion region,
an alternative to the region derived by Yang and Qin,
is characterized by entropy functions
for arbitrary general correlated sources.
Achievability is shown by constructing a code
based on constrained-random number generators.
\end{abstract}

\section{Introduction}

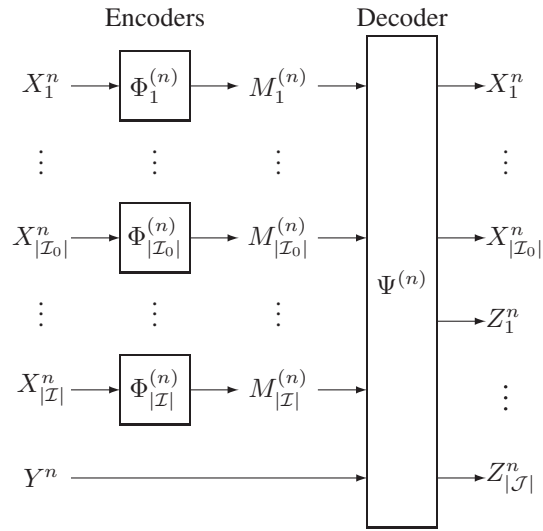
\begin{figure}
\begin{center}
 \unitlength 0.65mm
 \begin{picture}(120,115)(0,0)
  \put(10,95){\makebox(0,0){$X^n_1$}}
  \put(10,81){\makebox(0,0){$\vdots$}}
  \put(10,63){\makebox(0,0){$X^n_{|\I_0|}$}}
  \put(10,50){\makebox(0,0){$\vdots$}}
  \put(10,33){\makebox(0,0){$X_{|\I|}^n$}}
  \put(10,15){\makebox(0,0){$Y^n$}}
  \put(16,95){\vector(1,0){10}}
  \put(16,64){\vector(1,0){10}}
  \put(16,33){\vector(1,0){10}}
  \put(16,15){\vector(1,0){60}}
  \put(33,109){\makebox(0,0){Encoders}}
  \put(26,88){\framebox(14,14){$\Phi^{(n)}_1$}}
  \put(33,81){\makebox(0,0){$\vdots$}}
  \put(26,57){\framebox(14,14){$\Phi^{(n)}_{|\I_0|}$}}
  \put(33,50){\makebox(0,0){$\vdots$}}
  \put(26,26){\framebox(14,14){$\Phi^{(n)}_{|\I|}$}}
  \put(40,95){\vector(1,0){10}}
  \put(40,64){\vector(1,0){10}}
  \put(40,33){\vector(1,0){10}}
  \put(58,95){\makebox(0,0){$M^{(n)}_1$}}
  \put(58,81){\makebox(0,0){$\vdots$}}
  \put(58,64){\makebox(0,0){$M^{(n)}_{|\I_0|}$}}
  \put(58,50){\makebox(0,0){$\vdots$}}
  \put(58,33){\makebox(0,0){$M^{(n)}_{|\I|}$}}
  \put(66,95){\vector(1,0){10}}
  \put(66,64){\vector(1,0){10}}
  \put(66,33){\vector(1,0){10}}
  \put(83,109){\makebox(0,0){Decoder}}
  \put(76,5){\framebox(14,100){$\Psi^{(n)}$}}
  \put(90,95){\vector(1,0){10}}
  \put(90,64){\vector(1,0){10}}
  \put(100,95){\makebox(0,0)[l]{$X^n_1$}}
  \put(104,81){\makebox(0,0){$\vdots$}}
  \put(100,63){\makebox(0,0)[l]{$X^n_{|\I_0|}$}}
  \put(90,47){\vector(1,0){10}}
  \put(104,33){\makebox(0,0){$\vdots$}}
  \put(90,15){\vector(1,0){10}}
  \put(100,47){\makebox(0,0)[l]{$Z^n_1$}}
  \put(104,33){\makebox(0,0){$\vdots$}}
  \put(100,15){\makebox(0,0)[l]{$Z^n_{|\J|}$}}
 \end{picture}
\end{center}
\caption{
 Distributed Source Coding:
 $\I$ is the index set of sources and encoders,
 $\I_0$ is the index set of sources reproduced losslessly
 and $\J$ is the index set of other (possibly lossy) reproductions.
}
\label{fig:dsc}
\end{figure}

This paper investigates a general distributed source coding problem
on a multiple-access channel (Fig.~\ref{fig:dsc}),
which is formulated in \cite{JB08}.
The problem includes distributed lossless source coding \cite{SW73},
lossy source coding with (non-causal) side information
at the decoder \cite{WZ76},
lossless source coding
with coded side information \cite{AK75,KM79,GP79,HK80,W75},
and distributed lossy source coding \cite{B78,T78}.

The contributions of this paper are listed below:
\begin{itemize}
 \item 
 The multi-letter rate-distortion region, which is an alternative
 to the region derived in \cite{YQ06a,YQ06b},
 is characterized by entropy functions
 for arbitrary general correlated sources,
 where the region is simplified
 by introducing losslessly reproduced sources.
 \item
 It is shown that the multi-letter rate-distortion region
 is achievable with a code based on constrained-random number
 generators \cite{CRNG,CRNGVLOSSY}.
 It provides a construction alternative to that described in \cite{YQ06a}.
\end{itemize}

This paper is organized as follows.
Some definitions and notations are introduced
in Section \ref{sec:definition}.
The definition of the rate-distortion region
is given in Section \ref{sec:rate-distortion-region},
where a comparison with previous results is described.
The proof of the converse is given in Section \ref{sec:converse}.
Code construction is introduced in Section \ref{sec:construction}
and the proof of achievability is given in Section \ref{sec:proof-crng}.

\section{Definitions and Notations}
\label{sec:definition}

Sets are written in calligraphic style (e.g. $\U$).
If $\U$ is a set and $\V_u$ is also a set for each $u\in\U$,
we use the notation $\V_{\U}\equiv\Prod_{u\in\U}\V_u$.
We use the notation $v_{\U}\equiv\{v_u\}_{u\in\U}$
to represent the set of elements
(e.g. sequences, random variables, functions) $v_u$ with index $u\in\U$.
We use the notation $|\U$| to represent the cardinality of $\U$.
We use the notation $2^{\U}\setminus\{\emptyset\}$
to represent the family of all non-empty subsets of $\U$.

A random variable and its realization
are denoted in roman (e.g. $U$ and $u$),
where the range of the random variable is written
in corresponding calligraphic style (e.g. $\U$).
For a given $n\in\NN\equiv\{1,2,\ldots\}$,
an $n$-dimensional vector random variable
is denoted by superscript $n$ (e.g. $U^n$),
where the range of the random variable is written in corresponding
calligraphic style (e.g. $\U^n$).
A $n$-dimensional vector (the realization of a vector random variable)
is denoted by superscript $n$ or in boldface (e.g. $u^n$ or $\uu$).

Let $\chi$ be the support function defined as
\begin{gather*}
 \chi(\mathrm{statement})
 \equiv
 \begin{cases}
  1
  &\text{if the $\mathrm{statement}$ is true}
  \\
  0
  &\text{if the $\mathrm{statement}$ is false}.
 \end{cases}
\end{gather*}
Throughout this paper,
we assume that the sequence of correlated random variables
$(\WW_{\I},\XX_{\I},\YY,\ZZ_{\J})\equiv
\{(W^n_{\I},X^n_{\I},Y^n,Z^n_{\J})\}_{n=1}^{\infty}$
is a joint general source,
where we do not assume conditions such as consistency,
stationarity, and ergodicity.
We assume that the alphabets $\X^n_i$ and $\Y^n$ of $X_i^n$ and $Y^n$
are the Cartesian product of set $\X_i$ and $\Y$, respectively.
On the contrary,
the alphabets $\W^n_i$ and $\Z^n_j$ of $W^n_i$ and $Z^n_j$
are not restricted to the Cartesian product
of set $\W_i$ and $\Z_j$, respectively.
We use the notations $\W^n_i$ and $\Z^n_j$ to make it
easier to understand the correspondence
with the stationary memoryless case.
We also assume that $\W^n_i$ is a finite set
but $\X_i^n$, $\Y^n$, and $\Z^n_j$
are allowed to be infinite sets under appropriate conditions,
where the summations are replaced with integrals.
We use the information-spectrum methods \cite{HAN,HV93}
summarized in Appendix \ref{sec:ispec}.

\section{Definition of Rate-Distortion Region}
\label{sec:rate-distortion-region}

Let $\I$ be the index set of sources and encoders.
Let $\I_0$ be the index set of sources which are reproduced losslessly,
where $\I_0\subset\I$.
Let $\J$ be the index set of other reproductions,
where we assume that $\I_0\cap\J=\emptyset$.
Let $n\in\NN$ denote the block length.

Let $\XX_{\I}\equiv\{\XX_i\}_{i\in\I}$
be a set of correlated general sources,
where $\XX_i\equiv\{X^n_i\}_{n=1}^{\infty}$.
Let $\YY\equiv\{Y^n\}_{n=1}^{\infty}$,
be (non-causal) side information available only at the decoder
correlated with $\XX_{\I}$.
We assume that Encoder $i$ observes source $X^n_i$
and transmits the codeword $M^{(n)}_i\in\M^{(n)}_i$ to the decoder.
Let $\ZZ_{\J}\equiv\{\ZZ_j\}_{j\in\J}$ be a set of reproductions
other than $\XX_{\I_0}$,
where $\ZZ_j\equiv\{Z^n_j\}_{n=1}^{\infty}$.
We assume that the decoder reproduces $(X^n_{\I_0},Z^n_{\J})$
after observing the set of codewords
$M^{(n)}_{\I}\equiv\{M^{(n)}_i\}_{i\in\I}$
and the uncoded side information $Y^n$.

First, we introduce an operational definition
of the rate-distortion region,
which is analogous to that defined in \cite{JB08}.
\begin{df}
Let $R_{\I}\equiv\{R_i\}_{i\in\I}$ and $D_{\J}\equiv\{D_j\}_{j\in\J}$
be the set of positive numbers.
Then a rate-distortion pair $(R_{\I},D_{\J})$ is {\em achievable}
for a given set of distortion measures
$\{
 d^{(n)}_j:\X^n_{\I}\times\Y^n\times\Z^n_j\to[0,\infty)
 \}_{j\in\J,n\in\NN}$
iff there is a sequence of codes
$\{
 (\{\vphi^{(n)}_i\}_{i\in\I},\{\psi^{(n)}_j\}_{j\in\I_0\cup\J})
 \}_{n=1}^{\infty}$
consisting of
encoding functions $\vphi^{(n)}_i:\X_i^n\to\M_i^{(n)}$
and reproducing functions
$\psi^{(n)}_j:\M^{(n)}_{\I}\times\Y^n\to\Z^{(n)}_j$
satisfying
\begin{gather}
 \limsupn \frac {\log|\M^{(n)}_i|}n \leq R_i
 \ \text{for all $i\in\I$}
 \label{eq:rate}
 \\
 \limn\Prob\lrsb{
  X^n_i\neq Z_i^n
 }=0
 \ \text{for all $i\in\I_0$}
 \label{eq:lossless}
 \\
 \limn
 \Prob\lrsb{
  d^{(n)}_j(X_{\I}^n,Y^n,Z_j^n)
  > D_j+\delta
 }=0
 \ \text{for all $j\in\J$ and $\delta>0$},
 \label{eq:lossy}
\end{gather}
where $\M^{(n)}_i$ is a finite set for all $i\in\I$
and $Z^n_j\equiv\psi_j^{(n)}(\{\vphi^{(n)}_i(X_i^n)\}_{i\in\I},Y^n)$
is the $j$-th reproduction for each $j\in\I_0\cup\J$.
The {\em rate-distortion region} $\ROP$
under the maximum-distortion criterion is defined as
the closure of the set of all achievable rate-distortion pairs.
\end{df}

\begin{rem}
Although the condition (\ref{eq:lossless})
can be included in (\ref{eq:lossy})
by letting $d^{(n)}_i(\xx_i,\yy,\zz_i)\equiv\chi(\xx_i=\zz_i)$
and $D_i\equiv0$ for each $i\in\I_0$,
we separate the conditions and index sets $\I_0$ and $\J$
to simplify the characterization of the rate-distortion region.
It should also be noted that
side information $Y^n$ can be included in the target sources
by assuming that $Y^n$ is encoded and decoded with infinite rate.
\end{rem}

Next, let us define region $\RIT$ as follows.
Let $\Ipc$ be defined as
\begin{align*}
 \Ipc
 &\equiv\I\setminus\I'.
\end{align*}
Let $\od_j(\XX_{\I},\YY,\ZZ_j)$ be defined as
\begin{equation*}
 \od_j(\XX_{\I},\YY,\ZZ_j)
 \equiv
 \plimsupn d^{(n)}_j(X^n_{\I},Y^n,Z^n_j)
\end{equation*}
for each $j\in\J$.
\begin{df}
\label{df:RIT}
Let $(\WW_{\I\setminus\I_0},\ZZ_{\J})$ be a set of general sources,
where $\WW_i\equiv \{W^n_i\}_{n=1}^{\infty}$ for each $i\in\I\setminus\I_0$
and $\ZZ_j\equiv \{Z^n_j\}_{n=1}^{\infty}$ for each $j\in\J$.
Let $\RIT(\WW_{\I\setminus\I_0},\ZZ_{\J})$
be defined as the set of all $(R_{\I},D_{\J})$ satisfying
\begin{align}
 \sum_{i\in\I'\cap\I_0}R_i
 &\geq
 \oH(\XX_{\I'\cap\I_0}|\WW_{\I\setminus\I_0},\XX_{\Ipc\cap\I_0},\YY)
 \label{eq:IT-RI'capI0}
 \\
 \sum_{i\in\I'\setminus\I_0}R_i
 &\geq
 \oH(\WW_{\I'\setminus\I_0}|\WW_{\Ipc\setminus\I_0},\YY)
 -
 \sum_{i\in\I'\setminus\I_0}
 \uH(\WW_i|\XX_i)
 \label{eq:IT-RI'setminusI0}
 \\
 D_j
 &\geq
 \od_j(\XX_{\I},\YY,\ZZ_j)
 \label{eq:IT-D}
\end{align}
for all $\I'\in2^{\I}\setminus\{\emptyset\}$ and $j\in\J$. 
Region $\RIT$ is defined by the union of
$\RIT(\WW_{\I\setminus\I_0},\ZZ_{\J})$
over a pair of general sources $(\WW_{\I\setminus\I_0},\ZZ_{\J})$ 
where the following conditions are satisfied:
\begin{gather}
 (W^n_{\I\setminus[\I_0\cup\{i\}]},X^n_{\I\setminus\{i\}},Y^n)
 \markov
 X^n_i
 \markov
 W^n_i
 \label{eq:IT-markov-encoder}
 \\
 X^n_{\I\setminus\I_0}
 \markov
 (W^n_{\I\setminus\I_0},X^n_{\I_0},Y^n)
 \markov
 Z^n_{\J}
\label{eq:IT-markov-decoder}
\end{gather}
for all $i\in\I\setminus\I_0$ and $n\in\NN$.
Optionally, $Z^n_{\J}$ is allowed to be restricted to
being the deterministic function
of $(W^n_{\I\setminus\I_0},X^n_{\I_0},Y^n)$.
\end{df}

\begin{rem}
We can introduce auxiliary real-valued variables 
$\{r_i\}_{i\in\I\setminus\I_0}$
to obtain the bounds
\begin{align}
 0
 \leq
 r_i
 &\leq
 \uH(\WW_i|\XX_i)
 \label{eq:RIT-ri}
 \\
 \sum_{i\in\I'\cap\I_0}R_i
 &\geq
 \oH(\XX_{\I'\cap\I_0}|\WW_{\I\setminus\I_0},\XX_{\Ipc\cap\I_0},\YY)
 \label{eq:RIT-RI'capI_0}
 \\
 \sum_{i\in\I'\setminus\I_0}[r_i+R_i]
 &\geq
 \oH(\WW_{\I'\setminus\I_0}|\WW_{\Ipc\setminus\I_0},\YY)
 \label{eq:RIT-[r+R]I'setminusI_0}
\end{align}
for all $i\in\I\setminus\I_0$ and $\I'\in2^{\I}\setminus\{\emptyset\}$.
By using the Fourier-Motzkin method \cite[Appendix E]{EK11}
to eliminate $\{r_i\}_{i\in\I\setminus\I_0}$,
we have the fact that they are equivalent to
(\ref{eq:IT-RI'capI0}) and (\ref{eq:IT-RI'setminusI0})
for all $\I'\in2^{\I}\setminus\{\emptyset\}$.
\end{rem}

Furthermore, let us define region $\RCRNG$ as follows.
\begin{df}
\label{df:RCRNG}
Let $(\WW_{\I\setminus\I_0},\ZZ_{\J})$ be a set of general sources,
where $\WW_i\equiv \{W^n_i\}_{n=1}^{\infty}$ for each $i\in\I\setminus\I_0$
and $\ZZ_j\equiv \{Z^n_j\}_{n=1}^{\infty}$ for each $j\in\J$.
Let $\RCRNG(\WW_{\I\setminus\I_0},\ZZ_{\J})$
be defined as the set of $(R_{\I},D_{\J})$ where 
there are real-valued variables
$\{r_i\}_{i\in\I\setminus\I_0}$ satisfying
\begin{align}
 0
 \leq
 r_i
 &\leq
 \uH(\WW_i|\XX_i)
 \label{eq:RCRNG-r}
 \\
 \sum_{i\in\I'\setminus\I_0}[r_i+R_i]
 +
 \sum_{i\in\I'\cap\I_0}R_i
 &\geq
 \oH(\WW_{\I'\setminus\I_0},\XX_{\I'\cap\I_0}
  |\WW_{\Ipc\setminus\I_0},\XX_{\Ipc\cap\I_0},\YY)
 \label{eq:RCRNG-r+R}
\end{align}
and (\ref{eq:IT-D}) for all $i\in\I\setminus\I_0$,
$\I'\in2^{\I}\setminus\{\emptyset\}$, and $j\in\J$.
Then region $\RCRNG$ is defined by the union of
$\RCRNG(\WW_{\I\setminus\I_0},\ZZ_{\J})$
over a pair of general sources $(\WW_{\I\setminus\I_0},\ZZ_{\J})$
satisfying (\ref{eq:IT-markov-encoder}) and (\ref{eq:IT-markov-decoder})
for all $i\in\I$ and $n\in\NN$.
Optionally, $Z^n_{\J}$ is allowed to
be restricted to being the deterministic function
of $(W^n_{\I\setminus\I_0},X^n_{\I_0},Y^n)$.
\end{df}

\begin{rem}
By using the Fourier-Motzkin method \cite[Appendix E]{EK11},
we can eliminate auxiliary variables $\{r_i\}_{i\in\I\setminus\I_0}$
and obtain the bound
\begin{equation*}
 \sum_{i\in\I'}R_i
 \geq
 \oH(\WW_{\I'\setminus\I_0},\XX_{\I'\cap\I_0}
  |\WW_{\Ipc\setminus\I_0},\XX_{\Ipc\cap\I_0},\YY)
 -
 \sum_{i\in\I'\setminus\I_0}
 \uH(\WW_i|\XX_i)
\end{equation*}
for all $\I'\in2^{\I}\setminus\{\emptyset\}$,
which is equivalent to (\ref{eq:RCRNG-r}) and (\ref{eq:RCRNG-r+R})
for all $\I'\in2^{\I}\setminus\{\emptyset\}$.
We introduce the auxiliary variables $\{r_i\}_{i\in\I\setminus\I_0}$
in the definition of $\RCRNG$
because expression (\ref{eq:RCRNG-r}) and (\ref{eq:RCRNG-r+R})
are related directly to the proof of
both the converse and achievability.
It should be noted here that both
the righthand side of (\ref{eq:RCRNG-r}) and (\ref{eq:RCRNG-r+R})
are monomial entropy functions,
that is, they have explicit operational interpretations
as explained in Remark \ref{rem:interpretation}
in Section \ref{sec:construction}.
\end{rem}

In subsequent sections, we show the following theorem.
\begin{thm}
\label{thm:ROP=RIT=RCRNG}
For a set of general correlated sources $(\XX_{\I},\YY)$,
we have
\begin{equation*}
 \ROP=\RIT=\RCRNG.
\end{equation*}
\end{thm}
\begin{IEEEproof}
The proof of the theorem consists of the following three facts:
\begin{enumerate}
 \item
 the converse $\ROP\subset\RIT$,
 which is shown in Section \ref{sec:converse},
 where (\ref{eq:IT-RI'capI0}) and (\ref{eq:IT-RI'setminusI0})
 are replaced by equivalent conditions 
 (\ref{eq:RIT-ri})--(\ref{eq:RIT-[r+R]I'setminusI_0});
 \item
 achievability $\RCRNG\subset\ROP$,
 which is shown in
 Sections \ref{sec:construction} and \ref{sec:proof-crng};
 \item
 the relation $\RIT\subset\RCRNG$
 derived immediately from the fact that
 (\ref{eq:RIT-ri})--(\ref{eq:RIT-[r+R]I'setminusI_0}) implies
 \begin{align}
	\sum_{i\in\I'\setminus\I_0}
	[r_i+R_i]
	+
	\sum_{i\in\I'\cap\I_0}
	R_i
	&\geq
	\oH(\WW_{\I'\setminus\I_0}|\WW_{\Ipc\setminus\I_0},\YY)
	+
	\oH(\XX_{\I'\cap\I_0}|
	 \WW_{\I\setminus\I_0},\XX_{\Ipc\cap\I_0},\YY)
	\notag
	\\
	&\geq
	\oH(\WW_{\I'\setminus\I_0}|
	 \WW_{\Ipc\setminus\I_0},\XX_{\Ipc\cap\I_0},\YY)
	+
	\oH(\XX_{\I'\cap\I_0}|
	 \WW_{\I\setminus\I_0},\XX_{\Ipc\cap\I_0},\YY)
	\notag
	\\
	&\geq
	\oH(\WW_{\I'\setminus\I_0},\XX_{\I'\cap\I_0}|
	 \WW_{\Ipc\setminus\I_0},\XX_{\Ipc\cap\I_0},\YY)
 \end{align}
 for all 
 $\I'\in 2^{\I}\setminus\{\emptyset\}$,
 where the second inequality comes from
 Lemma \ref{lem:oH(U|V)>oH(U|VV')}	in Appendix~\ref{sec:ispec},
 and the third inequality comes from
 Lemma \ref{lem:oH(UU'|V)<oH(U'|UV)+oH(U|V)} in Appendix~\ref{sec:ispec}.
\end{enumerate}
\end{IEEEproof}

The following examples are particular cases of the distributed
source coding problem.
In the examples, we discuss only the achievable regions,
which are actually optimal regions,
by specifying auxiliary random variables.
It should be noted that
the converse part can be shown from past studies.

\begin{example}
When $\I=\I_0$ and $\J=\emptyset$,
the rate-distortion region represents
the distributed lossless source coding region.
Since $\I'\setminus\I_0=\emptyset$
for all $\I'\in2^{\I}\setminus\{\emptyset\}$,
we have
\begin{align*}
 \sum_{i\in\I'}
 R_i
 &\geq
 \oH(\XX_{\I'}|\XX_{\Ipc},\YY)
\end{align*}
for all $\I'\in 2^{\I}\setminus\{\emptyset\}$ from (\ref{eq:IT-RI'capI0}).
This region is given in \cite{MK95}
for the case where $\I=\I_0=\{1,2\}$ and $Y^n$ is a constant.
In particular,
when $\I=\I_0=\{1\}$,
the rate-distortion region represents
the point-to-point lossless source coding region introduced in \cite{HV93}
as
\begin{align*}
 R_1
 &\geq
 \oH(\XX_1).
\end{align*}
It should be noted that the general region
for the case of two or more decoders is given in \cite{CRNG-MULTI}.
\end{example}

\begin{example}
When $\I=\J=\{1\}$, $\I_0=\emptyset$, and $Y^n$ is a constant,
the rate-distortion region
with a given distortion measure
$d^{(n)}_1:\X^n_1\times\Z^n_1\to[0,\infty)$
and distortion level $D_1\in[0,\infty)$
represents the point-to-point rate-distortion region
under the maximum-distortion criterion introduced in \cite{CRNG,SV93}.
By letting $\ZZ_1=\WW_1$,
we have
\begin{align*}
 R_1
 &\geq
 \oH(\WW_1)
 -\uH(\WW_1|\XX_1)
 \\
 D_1
 &\geq
 \od_1(\XX_1,\WW_1)
\end{align*}
from (\ref{eq:IT-RI'setminusI0}) and (\ref{eq:IT-D}).
It is shown in \cite{CRNG} that
this region is equivalent to the region specified by
\begin{align*}
 R_1
 &\geq
 \oI(\XX_1;\WW_1)
 \\
 D_1
 &\geq
 \od_1(\XX_1,\WW_1)
\end{align*}
given in \cite{SV93}.
\end{example}

\begin{example}
When $\I=\J=\{1\}$ and $\I_0=\emptyset$,
the rate-distortion region
with given distortion measure $d^{(n)}_1:\X_1^n\times\Z_1^n\to[0,\infty)$
and distortion level $D_1\in[0,\infty)$
represents the region
of lossy source coding with non-causal side information at the decoder
introduced in \cite{IM02}.
By letting $\ZZ_1=\WW_1$,
we have
\begin{align*}
 R_1
 &\geq
 \oH(\WW_1|\YY)
 -\uH(\WW_1|\XX_1)
 \\
 D_1
 &\geq
 \od_1(\XX_1,\WW_1)
 \notag
\end{align*}
from (\ref{eq:IT-RI'setminusI0}) and (\ref{eq:IT-D}).
The equivalence to the region
\begin{align*}
 R_1
 &\geq
 \oI(\WW_1;\XX_1)
 -\uI(\WW_1;\YY)
 \\
 D_1
 &\geq
 \od_1(\XX_1,\WW_1)
 \notag
\end{align*}
introduced in \cite{IM02} 
can be shown by using the achievability and converse
of the two regions.
\end{example}

\begin{example}
When $\I\setminus\I_0=\{\helper\}$, $\J=\emptyset$,
and $Y^n$ is a constant,
the rate-distortion region represents the region
of lossless source coding with a helper
that provides the coded side information.
We have the achievable region specified by
\begin{align*}
 R_{\helper}
 &\geq
 \oH(\WW_{\helper})
 -
 \uH(\WW_{\helper}|\XX_{\helper})
 \\
 \sum_{i\in\I'}
 R_i
 &\geq
 \oH(\XX_{\I'}|\WW_{\helper},\XX_{\I_0\setminus\I'})
\end{align*}
for all $\I'\subset2^{\I_0}\setminus\{\emptyset\}$
from (\ref{eq:IT-RI'capI0}) and (\ref{eq:IT-RI'setminusI0}).
This region is equivalent to the region specified by
\begin{align*}
 R_{\helper}
 &\geq
 \oH(\WW_{\helper}|\XX_{\I_0})
 -
 \uH(\WW_{\helper}|\XX_{\helper})
 \\
 \sum_{i\in\I'}
 R_i
 &\geq
 \oH(\XX_{\I'}|\WW_{\helper},\XX_{\I_0\setminus\I'})
 \\
 R_{\helper}
 +
 \sum_{i\in\I'}
 R_i
 &\geq
 \oH(\WW_{\helper},\XX_{\I'}|\XX_{\I_0\setminus\I'})
\end{align*}
for all $\I'\subset2^{\I_0}\setminus\{\emptyset\}$.
These regions are equivalent to
the region specified by
\begin{align*}
 R_{\helper}
 &\geq
 \oI(\XX_{\helper};\WW_{\helper})
 \\
 \sum_{i\in\I'}
 R_i
 &\geq
 \oH(\XX_{\I'}|\WW_{\helper},\XX_{\I_0\setminus\I'}),
\end{align*}
which is the extension of that derived in \cite{MK95}.
The equivalence can be shown by using the achievability
and converse of these regions.
\end{example}

\begin{example}
When $\I=\J$ and $\I_0=\emptyset$,
the rate-distortion region
with given distortion measure $d^{(n)}_i:\X_i^n\times\Z_i^n\to[0,\infty)$
and distortion levels $D_i\in[0,\infty)$
represents the region of distributed lossy source coding,
where the conditions (\ref{eq:IT-RI'capI0})--(\ref{eq:IT-D}) 
are reduced to
\begin{align*}
 \sum_{i\in\I'}
 R_i
 &\geq
 \oH(\WW_{\I'}|\WW_{\Ipc},\YY)
 -
 \sum_{i\in\I'}
 \uH(\WW_i|\XX_i)
 \\
 D_i
 &\geq
 \od_i(\XX_i,\ZZ_i)
\end{align*}
for all $\I'\in2^{\I}\setminus\{\emptyset\}$ and $i\in\I$.
This is a rate-distortion region
alternative to that derived in \cite{YQ06a,YQ06b}.
\end{example}

\begin{example}
Let $\WW_i\equiv\XX_i$ for each $i\in\I_0$.
Let $d_j$ be the bounded single letter distortion measure
and
\begin{equation*}
 d_j^{(n)}(\xx_{\I},\yy,\zz_j)\equiv
 \frac1n\sum_{k=1}^nd_j(x_{\I,k},y_k,z_{j,k})
\end{equation*}
for each $j\in\J$.
When $(\WW_{\I},\XX_{\I},\ZZ_{\I})$ is stationary memoryless,
we can replace
(\ref{eq:IT-D}) by
\begin{equation*}
 D_j\geq E_{X_{\I}YZ_j}[d_j(X_{\I},Y,Z_j)]
\end{equation*}
from the law of large numbers.
Since $H(W_i|X_i)=0$ for all $i\in\I_0$,
conditions (\ref{eq:IT-RI'capI0}) and (\ref{eq:IT-RI'setminusI0})
for all $\I'\in2^{\I}\setminus\{\emptyset\}$
are reduced to the region introduced in \cite{JB08} as
\begin{align}
 \sum_{i\in\I'}R_i
 &\geq
 H(W_{\I'}|W_{\Ipc},Y)
 -
 \sum_{i\in\I'}
 H(W_i|X_i)
 \notag
 \\
 &=
 H(W_{\I'}|W_{\Ipc})
 -
 H(W_{\I'}|W_{\Ipc},X_{\I'},Y)
 \notag
 \\
 &=
 I(X_{\I'};W_{\I'}|W_{\Ipc},Y)
\end{align}
for all $\I'\in2^{\I}\setminus\{\emptyset\}$,
where the first equality 
comes from (\ref{eq:IT-markov-encoder}) with $n=1$.
With the additional condition that
$\I_0=\emptyset$ and $Y^n$ is a constant,
region $\RIT$
is equal to the multiple extension of the
Berger-Tung single-letter inner region \cite{B78,T78}
without time-sharing variable.
It should be noted here that
the Berger-Tung single-letter inner region 
is sub-optimal for particular cases \cite{SP21,WKA11}.
We could conclude that
the sub-optimality is caused by
restricting $(\WW_{\I},\XX_{\I},\YY,\ZZ_{\J})$
to being stationary memoryless,
where it has been reported that
the Berger-Tung single-letter inner region
can be improved by considering multi-letter extensions \cite{SP21}.
\end{example}

\section{Proof of Converse}
\label{sec:converse}
This section argues the converse part $\ROP\subset\RIT$.
It should be noted here that the introduction of
auxiliary variables $\{r_i\}_{i\in\I}$ simplifies the proof.

Let us assume that $(R_{\I},D_{\J})\in\ROP$.
Then we have the fact that 
there is a sequence  $\{(\vphi^{(n)}_{\I},\psi^{(n)}_{\J})\}_{n=1}^{\infty}$
satisfying (\ref{eq:rate})--(\ref{eq:lossy}).
Let
\begin{align}
 W^n_i
 &\equiv
 \vphi^{(n)}_i(X^n_i)
 \label{eq:converse-W}
 \\
 Z^n_j
 &\equiv
 \psi^{(n)}_j(W^n_{\I},Y^n).
\end{align}
Then we have the Markov chains
(\ref{eq:IT-markov-encoder})
and (\ref{eq:IT-markov-decoder}).
From the definition of $\plimsupn$ in Appendix \ref{sec:ispec},
we have the fact that (\ref{eq:lossy}) is equivalent to
\begin{equation*}
 \plimsupn d^{(n)}_j(X^n_{\I},Y^n,Z^n_j)\leq D_j
\end{equation*}
and (\ref{eq:IT-D}) for all $j\in\J$.

Let $r_i\equiv 0$ for each $i\in\I\setminus\I_0$.
Then, from Lemma \ref{lem:oH>uH>0} in Appendix \ref{sec:ispec}, we have
\begin{align*}
 0
 \leq
 r_i
 &\leq
 \uH(\WW_i|\XX_i)
\end{align*}
for all $i\in\I\setminus\I_0$.
From (\ref{eq:rate}), we have
\begin{align}
 R_i
 &\geq
 \limsupn \frac{\log|\M^{(n)}_i|}n
 \notag
 \\
 &\geq
 \oH(\WW_i)
 \label{eq:converse-Ri}
\end{align}
for all $i\in\I$,
where the second inequality comes from
Lemma \ref{lem:bound-by-cardinality} in Appendix \ref{sec:ispec}.
Then we have 
\begin{align}
 \sum_{i\in\I'\setminus\I_0}
 [r_i+R_i]
 &=
 \sum_{i\in\I'\setminus\I_0}
 R_i
 \notag
 \\
 &\geq
 \sum_{i\in\I'\setminus\I_0}
 \oH(\WW_i)
 \notag
 \\
 &\geq
 \oH(\WW_{\I'\setminus\I_0})
 \notag
 \\
 &\geq
 \oH(\WW_{\I'\setminus\I_0}|\WW_{\Ipc\setminus\I_0},\YY)
 \label{eq:converse-[r+R]I'setminusI0}
\end{align}
for all $\I'\in2^{\I}\setminus\{\emptyset\}$,
where the first inequality comes from (\ref{eq:converse-Ri}),
the second inequality comes from
Lemmas \ref{lem:oH(UU'|V)<oH(U'|UV)+oH(U|V)}
and \ref{lem:oH(U|V)>oH(U|VV')}
in Appendix \ref{sec:ispec},
and the last inequality comes from
Lemma \ref{lem:oH(U|V)>oH(U|VV')}
in Appendix \ref{sec:ispec}.

Furthermore, we have
\begin{align}
 \sum_{i\in\I'\cap\I_0}
 R_i
 &\geq
 \sum_{i\in\I'\cap\I_0}
 \oH(\WW_i)
 \notag
 \\
 &\geq
 \oH(\WW_{\I'\cap\I_0})
 \notag
 \\
 &\geq
 \oH(\WW_{\I'\cap\I_0}
	|\WW_{[\I'\cap\I_0]^{\complement}},\XX_{\Ipc\cap\I_0},\YY)
 \notag
 \\
 &\geq
 \oH(\XX_{\I'\cap\I_0}
	|\WW_{[\I'\cap\I_0]^{\complement}},\XX_{\Ipc\cap\I_0},\YY)
 \notag
 \\
 &\geq
 \oH(\WW_{\Ipc\cap\I_0},\XX_{\I'\cap\I_0}
	|\WW_{\I\setminus\I_0},\XX_{\Ipc\cap\I_0},\YY)
 -
 \oH(\WW_{\Ipc\cap\I_0}
	|\WW_{\I\setminus\I_0},\XX_{\Ipc\cap\I_0},\YY)
 \notag
 \\
 &=
 \oH(\WW_{\Ipc\cap\I_0},\XX_{\I'\cap\I_0}
	|\WW_{\I\setminus\I_0},\XX_{\Ipc\cap\I_0},\YY)
\notag
 \\
 &\geq
 \oH(\XX_{\I'\cap\I_0}
	|\WW_{\I\setminus\I_0},\XX_{\Ipc\cap\I_0},\YY)
 \label{eq:converse-RI'capI0}
\end{align}
for all $\I'\in2^{\I}\setminus\{\emptyset\}$,
where the first inequality comes from (\ref{eq:converse-Ri}),
the second inequality comes from
Lemmas \ref{lem:oH(UU'|V)<oH(U'|UV)+oH(U|V)} and \ref{lem:oH(U|V)>oH(U|VV')}
in Appendix \ref{sec:ispec},
the third inequality comes from
Lemma \ref{lem:oH(U|V)>oH(U|VV')} in Appendix \ref{sec:ispec},
the fourth inequality is shown by applying
Lemma \ref{lem:sw-bound} in Appendix \ref{sec:ispec}
together with the fact that
(\ref{eq:lossless}) and Lemma~\ref{lem:fano} in Appendix \ref{sec:ispec}
imply
\begin{align}
 0
 \leq
 \oH(\XX_{\I'\cap\I_0}
	|\WW_{\I'\cap\I_0},
	(\WW_{[\I'\cap\I_0]^{\complement}},\XX_{\Ipc\cap\I_0},\YY))
 &\leq
 \oH(\XX_{\I'\cap\I_0}|\WW_{\I},\YY)
 \notag
 \\
 &=0,
\end{align}
the fifth inequality comes from 
Lemma \ref{lem:oH(UU'|V)<oH(U'|UV)+oH(U|V)} in Appendix \ref{sec:ispec}
and the fact that
$[\I'\cap\I_0]^{\complement}$ is the disjoint union of
$\Ipc\cap\I_0$ and $\I\setminus\I_0$,
the equality comes from (\ref{eq:converse-W})
and Lemma \ref{lem:fano} in Appendix~\ref{sec:ispec},
and the last inequality comes from
Lemma \ref{lem:oH(UU'|V)>oH(U|V)} in Appendix~\ref{sec:ispec}.
Thus, we have the fact that $(R_{\I},D_{\J})\in\RIT$.
\hfill\IEEEQED

\begin{rem}
We now show another converse of $\ROP\subset\RCRNG$
directly without using (\ref{eq:converse-[r+R]I'setminusI0})
and (\ref{eq:converse-RI'capI0}).
We have (\ref{eq:RCRNG-r+R}) as
\begin{align}
 \sum_{i\in\I'\cap\I_0}
 R_i
 +
 \sum_{i\in\I'\setminus\I_0}
 [r_i+R_i]
 &=
 \sum_{i\in\I'}
 R_i
 \notag
 \\
 &\geq
 \sum_{i\in\I'}
 \oH(\WW_i)
 \notag
 \\
 &\geq
 \oH(\WW_{\I'})
 \notag
 \\
 &\geq
 \oH(\WW_{\I'}|\WW_{\Ipc},\XX_{\Ipc\cap\I_0},\YY)
 \notag
 \\
 &\geq
 \oH(\WW_{\I'\setminus\I_0},\XX_{\I'\cap\I_0}
	|\WW_{\Ipc},\XX_{\Ipc\cap\I_0},\YY)
 \notag
 \\
 &\geq
 \oH(\WW_{\I'\setminus\I_0},\WW_{\Ipc\cap\I_0},\XX_{\I'\cap\I_0}
	|\WW_{\Ipc\setminus\I_0},\XX_{\Ipc\cap\I_0},\YY)
 -
 \oH(\WW_{\Ipc\cap\I_0}
	|\WW_{\Ipc\setminus\I_0},\XX_{\Ipc\cap\I_0},\YY)
 \notag
 \\
 &=
 \oH(\WW_{\I'\setminus\I_0},\WW_{\Ipc\cap\I_0},\XX_{\I'\cap\I_0}
	|\WW_{\Ipc\setminus\I_0},\XX_{\Ipc\cap\I_0},\YY)
 \notag
 \\
 &\geq
 \oH(\WW_{\I'\setminus\I_0},\XX_{\I'\cap\I_0}
	|\WW_{\Ipc\setminus\I_0},\XX_{\Ipc\cap\I_0},\YY)
 \label{eq:converse-RI'+rI'}
\end{align}
for all $\I'\in2^{\I}\setminus\{\emptyset\}$,
where the first equality comes from $r_i=0$ for all $i\in\I'\cap\I_0$.
the first inequality comes from (\ref{eq:converse-Ri}),
the second inequality comes from
Lemmas \ref{lem:oH(UU'|V)<oH(U'|UV)+oH(U|V)} and \ref{lem:oH(U|V)>oH(U|VV')}
in Appendix \ref{sec:ispec},
the third inequality comes from
Lemma \ref{lem:oH(U|V)>oH(U|VV')} in Appendix \ref{sec:ispec},
the fourth inequality is shown by applying
Lemma \ref{lem:sw-bound} in Appendix \ref{sec:ispec}
together with the fact that
(\ref{eq:lossless}) and Lemma \ref{lem:fano} in Appendix \ref{sec:ispec}
imply 
\begin{align}
 0
 \leq
 \oH(\WW_{\I'\setminus\I_0},\XX_{\I'\cap\I_0}
	|\WW_{\I'},(\WW_{\Ipc},\XX_{\Ipc\cap\I_0},\YY))
 &\leq
 \oH(\WW_{\I'\setminus\I_0},\XX_{\I'\cap\I_0}|\WW_{\I},\YY)
 \notag
 \\
 &=0,
\end{align}
the fifth inequality comes from
Lemma \ref{lem:oH(UU'|V)<oH(U'|UV)+oH(U|V)} in Appendix \ref{sec:ispec},
the second equality comes from (\ref{eq:converse-W})
and Lemma \ref{lem:fano} in Appendix~\ref{sec:ispec},
and the last inequality comes from
Lemma \ref{lem:oH(UU'|V)>oH(U|V)} in Appendix \ref{sec:ispec}.
\end{rem}

\section{Code Construction}
\label{sec:construction}

This section introduces a source code
based on an idea drawn from \cite{CRNG,CRNGVLOSSY,CoCoNuTS-LOSSY};
a similar idea is found in \cite{GGWK24,YAG12}.
It should be noted that when sources are stationary memoryless,
tractable approximation algorithms
for a constrained-random-number generator
summarized in~\cite{SDECODING} are available.

First, we consider the case where $\I_0=\emptyset$,
that is, no source is reproduced losslessly.

For each $i\in\I$, let us introduce set $\C^{(n)}_i$
and function $f_i:\W^n_i\to\C^{(n)}_i$,
where the dependence of $f_i$ on $n$ is omitted.
For each $i\in\I$, let us introduce set
$\M^{(n)}_i$ and function $g_i:\W^n_i\to\M^{(n)}_i$,
where the dependence of $g_i$ on $n$ is omitted.
We can use sparse matrices as functions $f_i$, $g_i$
by assuming that $\W_i^n$, $\C^{(n)}_i$, and $\M^{(n)}_i$ are linear spaces
on the same finite field.

We define here a constrained-random-number generator
used by Encoder $i$ to generate $\ww_i\in\W_i^n$.
For given $\xx_i$ and $\cc_i$,
let $\tW^n_i$ be a random variable corresponding to the distribution
\begin{align}
 \mu_{\tW^n_i|C^{(n)}_iX^n_i}(\ww_i|\cc_i,\xx_i)
 &\equiv
 \frac{
  \mu_{W^n_i|X^n_i}(\ww_i|\xx_i)
  \chi(
   f_i(\ww_i)=\cc_i
  )
 }{
  \sum_{\ww_i}
  \mu_{W^n_i|X^n_i}(\ww_i|\xx_i)
  \chi(
   f_i(\ww_i)=\cc_i
  )
 }.
 \label{eq:tWi}
\end{align}
Encoder $i$ generates $\ww_i$
by using this constrained-random-number generator.
Then we define stochastic encoding function
$\Phi^{(n)}_i:\X_i^n\to\M^{(n)}_i$ as
\begin{align*}
 \Phi^{(n)}_i(\xx_i)
 &\equiv
 g_i(\ww_i),
\end{align*}
where the encoder claims an error when
the numerator of the righthand side of (\ref{eq:tWi}) is zero.
By using the technique described in \cite[Section VI]{CRNG},
we can represent
\begin{equation*}
 \Phi^{(n)}_i(\xx_i)
 =
 \vphi^{(n)}_i(\xx_i,S_i)
\end{equation*}
by using the deterministic function $\vphi^{(n)}_i$
and randomness $S_i$,
which is independent of $(X^n_{\I},Y^n,S_{\I\setminus\{i\}})$.
In the following, let us define components of codeword $\mm_i$ as
\begin{align*}
 \mm_i
 &\equiv
 \Phi^{(n)}_i(\xx_i).
\end{align*}

We define a constrained-random-number generator used by the decoder.
Decoder generates $\hww_{\I}$
by using a constrained-random-number generator with a distribution
given as
\begin{align*}
 \mu_{\hW^n_{\I}|C^{(n)}_{\I}M^{(n)}_{\I}Y^n}
 (\hww_{\I}|\cc_{\I},\mm_{\I},\yy)
 &\equiv
 \frac{
  \mu_{W_{\I}^n|Y^n}(\hww_{\I}|\yy)
  \chi((f,g)_{\I}(\hww_{\I})=(\cc_{\I},\mm_{\I}))
 }{
  \sum_{\hww_{\I}}
  \mu_{W_{\I}^n|Y^n}(\hww_{\I}|\yy)
  \chi((f,g)_{\I}(\hww_{\I})=(\cc_{\I},\mm_{\I}))
 }
\end{align*}
for given vectors $\cc_{\I}$ and $\mm_{\I}$,
where
\begin{equation*}
 \mu_{W_{\I}^n|Y^n}(\ww_{\I}|\yy)
 \equiv
 \frac{
  \sum_{\xx_{\I}\in\X^n}
  \mu_{X^n_{\I}Y^n}(\xx_{\I},\yy)
  \prod_{i\in\I}
  \mu_{W^n_i|X^n_i}(\ww_i|\xx_i)}
 {
  \sum_{\xx_{\I}\in\X^n}
  \mu_{X^n_{\I}Y^n}(\xx_{\I},\yy)
 }.
\end{equation*}

It should be noted that
we can also use either the maximum a posteriori probability decoder
or the typical-set decoder
instead of using the constrained-random number generator.
We define the decoding function
$\Psi^{(n)}:\Prod_{i\in\I}\M^{(n)}_i\times\Y^n\to\Z^n_{\J}$
as
\begin{align*}
 \Psi^{(n)}(\mm_{\I},\yy)
 &\equiv
 \{\zeta^{(n)}_j(\hww_{\I},\yy)\}_{j\in\J}
\end{align*}
by using functions
$\{\zeta^{(n)}_j:\W^n_{\I}\times\Y^n\to\Z^n_j\}_{j\in\J}$.
By using the technique described in \cite[Section VI]{CRNG},
we can represent
\begin{equation*}
 \Psi^{(n)}_i(\mm_{\I},\yy)
 =
 \psi^{(n)}_i(\mm_{\I},\yy,S_{\mathrm d})
\end{equation*}
by using deterministic function $\psi^{(n)}_i$
and randomness $S_{\mathrm d}$,
which is independent of $(X^n_{\I},Y^n,S_{\I})$.

Let
\begin{align}
 r_i
 &\equiv
 \frac{\log|\C^{(n)}_i|}n
 \label{eq:ri}
 \\
 R_i
 &\equiv
 \frac{\log|\M^{(n)}_i|}n
 \label{eq:Ri}
\end{align}
for each $i\in\I$,
where $R_i$ represents the transmission rate of Encoder $i$.
Let $\bs$ be the output of the source $(S_{\I},S_{\mathrm d})$
used by the constrained-random number generators
and
$\hZ^n_j
\equiv
\psi_j^{(n)}(\{\vphi_i^{(n)}(X^n_i,\bs_i)\}_{i\in\I},Y^n,\bs_{\mathrm d})$.
For a given $\delta>0$, let $\Error(f_{\I},g_{\I},\cc_{\I},\bs)$ 
be the error probability defined as
\begin{equation*}
 \Error(f_{\I},g_{\I},\cc_{\I},\bs)
 \equiv
 \Prob\lrsb{
  d^{(n)}_j(X^n_{\I},Y^n,\hZ^n_j)
  > D_j+\delta
  \ \text{for some}\ j\in\J
 },
\end{equation*}
where $\bs\equiv(\bs_{\I},\bs_{\mathrm d})$.
We introduce the following theorem, 
which implies the achievability part $\RCRNG\subset\ROP$.
The proof is given in Section~\ref{sec:proof-crng}.

\begin{thm}
\label{thm:crng}
For given set of general correlated sources $(\XX_{\I},\YY)$
and rate-distortion pair $(R_{\I},D_{\J})$,
let us assume that general sources $\WW_{\I}$,
functions
$\{\{\zeta^{(n)}_j
	:\W^n_{\I}\times\Y^n\to\Z^n_j\}_{j\in\J}\}_{n=1}^{\infty}$,
and numbers
$\{(r_i,R_i,)\}_{i\in\I}$, and $\{D_j\}_{j\in\J}$
satisfy
\begin{align}
 0
 \leq
 r_i
 &<
 \uH(\WW_i|\XX_i)
 \label{eq:CRNG-ri}
 \\
 \sum_{i\in\I'}
 [R_i+r_i]
 &>
 \oH(\WW_{\I'}|\WW_{\Ipc},\YY)
 \label{eq:CRNG-sum[Ri+ri]}
\end{align}
for all $i\in\I$ and $\I'\in2^{\I}\setminus\{\emptyset\}$,
and
\begin{equation}
 \limn
 \Prob\lrsb{
  d^{(n)}_j(X^n_{\I},Y^n,Z^n_j)
  > D_j+\delta
 }
 =0
 \label{eq:crng-dXY>D}
\end{equation}
for all $j\in\J$,
where the joint distribution of $(W^n_{\I},X^n_{\I},Y^n,Z^n_{\J})$
is given by
\begin{equation}
 \mu_{W^n_{\I}X^n_{\I}Y^nZ^n_{\J}}(\ww_{\I},\xx_{\I},\yy,\zz_{\J})
 \equiv
 \lrB{
  \prod_{j\in\J}
  \chi(\zz_j=\zeta^{(n)}_j(\ww_{\I},\yy))
 }
 \lrB{
	\prod_{i\in\I}
	\mu_{W^n_i|X^n_i}(\ww_i|\xx_i)
 }
 \mu_{X^n_{\I}Y^n}(\xx_{\I},\yy),
 \label{eq:crng-markov}
\end{equation}
which is equivalent to the Markov conditions
(\ref{eq:IT-markov-encoder}) and (\ref{eq:IT-markov-decoder}).
Then for all $\delta>0$ and all sufficiently large $n$
there are functions (sparse matrices)
$f_{\I}\equiv\{f_i\}_{i\in\I}$ and $g_{\I}\equiv\{g_i\}_{i\in\I}$,
and vectors $\cc_{\I}\equiv\{\cc_i\}_{i\in\I}$
and $\bs\equiv(\bs_{\I},\bs_{\mathrm d})$
such that
$\Error(f_{\I},g_{\I},\cc_{\I},\bs)<\delta$.
\end{thm}

\begin{rem}
\label{rem:interpretation}
Here, let us explain the interpretation of conditions 
(\ref{eq:CRNG-ri}) and (\ref{eq:CRNG-sum[Ri+ri]}). 
From (\ref{eq:crng-markov}),
the righthand side of (\ref{eq:CRNG-ri})
can be replaced by $\uH(\WW_i|\XX_{\I})$.
Condition (\ref{eq:CRNG-ri})
represents the limit of the randomness
of source $\WW_i$, where the randomness
is independent of the given source $\XX_{\I}$.
Since the rate of $\cc_i$ is $r_i$ that satisfies (\ref{eq:CRNG-ri}),
encoders and the decoder can share
the constant vectors $\{\cc_i\}_{i\in\I}$,
which is generated independent of $\{\xx_i\}_{i\in\I}$.
Condition (\ref{eq:CRNG-sum[Ri+ri]})
represents the Slepian-Wolf region \cite{C75,SW73}
of the correlated sources $\WW_{\I}$ with decoder side information $\YY$,
where the encoding rate of the source $\WW_i$
is reduced by $r_i$.
\end{rem}

Next, we consider the case where $\I_0\neq\emptyset$,
that is, there are some sources reproduced without distortion.
To apply Theorem~\ref{thm:crng},
we assume that $\I_0\subset\J$ and use the following definitions
\begin{align*}
 W^n_i
 &\equiv
 X^n_i
 \\
 |\C_i^{(n)}|
 &\equiv
 1
 \\
 d_i^{(n)}(\xx_{\I},\yy,\zz_i)
 &\equiv
 \chi(\xx_i\neq \zz_i)
 \\
 \zeta_i^{(n)}(\ww_{\I})
 &\equiv
 \ww_i
\end{align*}
for each $i\in\I_0$ and $n\in\NN$.
Then we have
\begin{align*}
 \uH(\WW_i|\XX_i)
 &=0
 \\
 r_i
 &=0
 \\
 \tW^n_i
 &=
 X^n_i
 \\
 W^n_{\I}
 &=
 (W^n_{\I\setminus\I_0},X^n_{\I_0})
\end{align*}
for all $i\in\I$ and $n\in\NN$.
We also have
the fact that $C^{(n)}_i=f_i(W^n_i)$ is a constant
and (\ref{eq:crng-dXY>D}) is satisfied for all $j\in\I_0$.
From Theorem \ref{thm:crng}, we have the fact that
there is a code when
$\{r_i\}_{i\in\I\setminus\I_0}$ and $\{R_i\}_{i\in\I}$ satisfy
(\ref{eq:CRNG-ri}) for all $i\in\I$
and (\ref{eq:CRNG-sum[Ri+ri]})
for all $\I'\in 2^{\I}\setminus\{\emptyset\}$.
Since $r_i=0$ and $\uH(\WW_i|\XX_i)=0$ for all $i\in\I_0$,
conditions (\ref{eq:CRNG-ri}) and (\ref{eq:CRNG-sum[Ri+ri]})
are reduced to
\begin{align*}
 0
 \leq
 r_i
 &<
 \uH(\WW_i|\XX_i)
 \\
 \sum_{i\in\I'\cap\I_0}
 R_i
 +
 \sum_{i\in\I'\setminus\I_0}
 [R_i+r_i]
 &>
 \oH(\WW_{\I'\setminus\I_0},\XX_{\I'\cap\I_0}
	|\WW_{\Ipc\setminus\I_0},\XX_{\Ipc\cap\I_0},\YY)
\end{align*}
for all $i\in\I\setminus\I_0$ and $\I'\in2^{\I}\setminus\{\emptyset\}$,
which are equivalent to
\begin{align*}
 \sum_{i\in\I'}
 R_i
 &>
 \oH(\WW_{\I'\setminus\I_0},\XX_{\I'\cap\I_0}
	|\WW_{\Ipc\setminus\I_0},\XX_{\Ipc\cap\I_0},\YY)
 -
 \sum_{i\in\I'\setminus\I_0}
 \uH(\WW_i|\XX_i)
\end{align*}
by eliminating $\{r_i\}_{i\in\I\setminus\I_0}$
using the Fourier-Motzkin method \cite[Appendix E]{EK11}.
Thus, by taking the closure,
we have the achievability $\RCRNG\subset\ROP$.

\section{Proof of Theorem~\ref{thm:crng}}
\label{sec:proof-crng}

In this section,
we use the lemmas on the hash property introduced in 
\cite{CRNG,CRNGVLOSSY,HASH,HASH-BC,HASH-WTC,CRNG-MULTI,CRNG-CHANNEL};
a similar idea is found in \cite{YAG12}.
The definitions and lemmas used in the proof are introduced
in Appendices \ref{sec:crng-expectation}--\ref{sec:proof-crp}.
For given $i\in\I$, pair of functions $(f,g)_i\equiv(f_i,g_i)$,
and pair of vectors $(\cc_i,\mm_i)\in\C^{(n)}_i\times\M^{(n)}_i$
we define
\begin{align*}
 \fC_{f_i}(\cc_i)
 &\equiv
 \{\ww_i: f_i(\ww_i)=\cc_i\}
 \\
 \fC_{f_{\I}}(\cc_{\I})
 &\equiv
 \{\ww_{\I}: f_i(\ww_i)=\cc_i\ \text{for all}\ i\in\I\}
 \\
 \fC_{(f,g)_{\I}}(\cc_{\I},\mm_{\I})
 &\equiv
 \{\ww_{\I}: f_i(\ww_i)=\cc_i, g_i(\ww_i)=\mm_i\ \text{for all}\ i\in\I\}
\end{align*}
for
$\zz_{\I}\equiv\{\zz_i\}_{i\in\I}$,
$\cc_{\I}\equiv\{\cc_i\}_{i\in\I}$,
and $\mm_{\I}\equiv\{\mm_i\}_{i\in\I}$,
where $(f,g)_{\I}(\zz_{\I})\equiv\{(f_i(\zz_i),g_i(\zz_i))\}_{i\in\I}$.
In the following, we use the following definitions:
\begin{align*}
 \E(f_i,\cc_i)
 &\equiv
 \lrb{
  \xx_i:
  \mu_{W_i|X_i}(\fC_{f_i}(\cc_i)|\xx_i)=0
 }
 \\
 \E(f_{\I},\cc_{\I})
 &\equiv
 \lrb{
  \xx_{\I}:
  \xx_i\in\E(f_i,\cc_i)\ \text{for some}\ i\in\I
 }
 \\
 \E(D_{\J})
 &\equiv
 \lrb{
  (\xx_{\I},\yy,\zz_{\J}):
  d_j^{(n)}(\xx_{\I},\yy,\zz_j)>D_j+\delta
  \ \text{for some}\ j\in\J
 }
 \\
 \mu_{X_{\I}}
 (\xx_{\I})
 &\equiv
 \sum_{\yy}
 \mu_{X_{\I}Y}
 (\xx_{\I},\yy)
 \\
 \mu_{W_{\I}|X_{\I}}
 (\ww_{\I}|\xx_{\I})
 &\equiv
 \prod_{i\in\I}
 \mu_{W_i|X_i}(\ww_i|\xx_i)
 \\
 \mu_{W_{\I}X_{\I}Y}
 (\ww_{\I},\xx_{\I},\yy)
 &\equiv
 \mu_{W_{\I}|X_{\I}}(\ww_{\I}|\xx_{\I})
 \mu_{X_{\I}Y}(\xx_{\I},\yy)
 \\
 \mu_{\tW_{\I}|X_{\I}C_{\I}}
 (\ww_{\I}|\xx_{\I},\cc_{\I})
 &\equiv
 \prod_{i\in\I}
 \frac{
  \mu_{W_i|X_i}(\ww_i|\xx_i)
  \chi(\cc_i=f_i(\ww_i))
 }
 {\mu_{W_i|X_i}(\fC_{f_i}(\cc_i)|\xx_i)}
 \\
 \mu_{\hW_{\I}|C_{\I}M_{\I}Y}
 (\hww_{\I}|\cc_{\I},\mm_{\I},\yy)
 &\equiv
 \frac{
  \mu_{W_{\I}|Y}(\hww_{\I}|\yy)
  \chi(\cc_{\I}=f_{\I}(\hww_{\I}),\mm_{\I}=g_{\I}(\hww_{\I}))
 }
 {\mu_{W_{\I}|Y}(\fC_{(f,g)_{\I}}(\cc_{\I},\mm_{\I})|\yy)}
 \\
 \mu_{Z_{\J}|W_{\I}Y}(\zz_{\J}|\ww_{\I},\yy)
 &\equiv
 \prod_{j\in\J}
 \chi(\zz_j=\eta^{(n)}_j(\ww_{\I},\yy)),
\end{align*}
where we can also let $\mu_{Z_{\J}|W_{\I}Y}$ 
be an arbitrary probability distribution in the following proof.

We let $S\equiv(S_{\I},S_{\mathrm d})$
and assume that $S$ is independent of $\XX_{\I}$.
Since the expectation over random variable $S$
is the expectation of the random variable corresponding to
the constrained-random number generators,
shown in Appendix~\ref{sec:crng-expectation},
we have
\begin{align}
 &
 E_S\lrB{
  \Error(f_{\I},g_{\I},\cc_{\I},S)
 }
 \notag
 \\*
 &\leq
 \sum_{\xx_{\I}\in\E(f_{\I},\cc_{\I})}
 \mu_{X_{\I}}(\xx_{\I})
 \notag
 \\*
 &\quad
 +
 \sum_{\substack{
   \xx_{\I},\yy,\ww_{\I},\hww_{\I},\zz_{\J}:\\
   \xx_{\I}\notin\E(f_{\I},\cc_{\I})\\
   \ww_{\I}\in\fC_{f_{\I}}(\cc_{\I})\\
   (\xx_{\I},\yy,\zz_{\J})\in\E(D_{\J})
 }}
 \mu_{Z_{\J}|W_{\I}Y}
 (\zz_{\J}|\hww_{\I},\yy)
 \mu_{\hW_{\I}|C_{\I}M_{\I}Y}
 (\hww_{\I}|\cc_{\I},g_{\I}(\ww_{\I}),\yy)
 \mu_{\tW_{\I}|C_{\I}X_{\I}}
 (\ww_{\I}|\cc_{\I},\xx_{\I})
 \mu_{X_{\I}Y}(\xx_{\I},\yy)
 \notag
 \\
 &\leq
 \sum_{i\in\I}
 \sum_{\xx_i\in\E(f_i,\cc_i)}
 \mu_{X_i}(\xx_i)
 \notag
 \\
 &\quad
 +
 \sum_{\substack{
   \xx_{\I},\yy,\ww_{\I},\hww_{\I},\zz_{\J}:\\
   \xx_{\I}\notin\E(f_{\I},\cc_{\I})\\
   \ww_{\I}\in\fC_{f_{\I}}(\cc_{\I})\\
   (\xx_{\I},\yy,\zz_{\J})\in\E(D_{\J})
 }}
 \mu_{Z_{\J}|W_{\I}Y}
 (\zz_{\J}|\hww_{\I},\yy)
 \mu_{\hW_{\I}|C_{\I}M_{\I}Y}
 (\hww_{\I}|\cc_{\I},g_{\I}(\ww_{\I}),\yy)
 \notag
 \\*
 &\qquad\qquad\qquad\qquad\qquad\qquad\qquad\qquad\qquad\qquad
 \cdot
 \lrbar{
  \mu_{\tW_{\I}|C_{\I}X_{\I}}
  (\ww_{\I}|\cc_{\I},\xx_{\I})
  -
  \mu_{W_{\I}|X_{\I}}
  (\ww_{\I}|\xx_{\I})
  \prod_{i\in\I}|\im\F_i|
 }
 \mu_{X_{\I}Y}(\xx_{\I},\yy)
 \notag
 \\
 &\quad
 +
 \sum_{\substack{
   \xx_{\I},\yy,\ww_{\I},\hww_{\I},\zz_{\J}:\\
   \xx_{\I}\notin\E(f_{\I},\cc_{\I})\\
   \ww_{\I}\in\fC_{f_{\I}}(\cc_{\I})\\
   \hww_{\I}\neq\ww_{\I}
 }}
 \mu_{Z_{\J}|W_{\I}Y}
 (\zz_{\J}|\hww_{\I},\yy)
 \mu_{\hW_{\I}|C_{\I}M_{\I}Y}
 (\hww_{\I}|\cc_{\I},g_{\I}(\zz_{\I}),\yy)
 \mu_{W_{\I}|X_{\I}}
 (\ww_{\I}|\xx_{\I})
 \mu_{X_{\I}Y}(\xx_{\I},\yy)
 \prod_{i\in\I}|\im\F_i|
 \notag
 \\
 &\quad
 +
 \sum_{\substack{
   \xx_{\I},\yy,\ww_{\I},\hww_{\I},\zz_{\J}:\\
   \xx_{\I}\notin\E(f_{\I},\cc_{\I})\\
   \ww_{\I}\in\fC_{f_{\I}}(\cc_{\I})\\
   \hww_{\I}=\ww_{\I}\\
   (\xx_{\I},\yy,\zz_{\J})\in\E(D_{\J})
 }}
 \mu_{Z_{\J}|W_{\I}Y}
 (\zz_{\J}|\hww_{\I},\yy)
 \mu_{\hW_{\I}|C_{\I}M_{\I}Y}
 (\hww_{\I}|\cc_{\I},g_{\I}(\ww_{\I}),\yy)
 \mu_{W_{\I}|X_{\I}}
 (\ww_{\I}|\xx_{\I})
 \mu_{X_{\I}Y}(\xx_{\I},\yy)
 \prod_{i\in\I}|\im\F_i|,
  \label{eq:proof-crng-dlc-error}
\end{align}
where,
in the first inequality,
we use the fact that
$\ww_{\I}\notin\fC_{f_{\I}}(\cc_{\I})$
implies
$\mu_{\tW_{\I}|C_{\I}X_{\I}}(\ww_{\I}|\cc_{\I},\xx_{\I})=0$.

The first term on the righthand side of (\ref{eq:proof-crng-dlc-error})
is evaluated as
\begin{align*}
 E_{C_{\I}}\lrB{
  \text{the first term of (\ref{eq:proof-crng-dlc-error})}
 }
 &=
 \sum_{i\in\I}
 \sum_{\substack{
   \xx_i,\cc_i:\\
   \xx_i\in\E(f_i,\cc_i)
 }}
 \frac{\mu_{X_i}(\xx_i)}{|\im\F_i|}.
\end{align*}

By letting $\I\equiv\{1,2,\ldots,|\I|\}$,
the second term on the righthand side of (\ref{eq:proof-crng-dlc-error})
is evaluated as
\begin{align}
 &
 E_{C_{\I}}\lrB{
  \text{the second term of (\ref{eq:proof-crng-dlc-error})}
 }
 \notag
 \\*
 &\leq
 \sum_{\substack{
	 \xx_{\I},\ww_{\I},\cc_{\I}:\\
	 \xx_{\I}\notin\E(f_{\I},\cc_{\I})\\
   \ww_{\I}\in\fC_{f_{\I}}(\cc_{\I})
 }}
 \lrbar{
  \frac{
   \mu_{\tW_{\I}|C_{\I}X_{\I}}
   (\ww_{\I}|\cc_{\I},\xx_{\I})
  }{\prod_{i\in\I}|\im\F_i|}
  -
  \mu_{W_{\I}|X_{\I}}
  (\ww_{\I}|\xx_{\I})
 }
 \mu_{X_{\I}}(\xx_{\I})
 \notag
 \\
 &=
 \sum_{\substack{
   \xx_{\I},\ww_{\I},\cc_{\I}:\\
   \xx_{\I}\notin\E(f_{\I},\cc_{\I})\\
   \ww_{\I}\in\fC_{f_{\I}}(\cc_{\I})\\
 }}
 \lrbar{
  \prod_{i\in\I}
  \frac{
   1
  }{
   \mu_{W_i|X_i}(\fC_{f_i}(\cc_i)|\xx_i)|\im\F_i|
  }
  -
  1
 }
 \mu_{W_{\I}|X_{\I}}
 (\ww_{\I}|\xx_{\I})
 \mu_{X_{\I}}(\xx_{\I})
 \notag
 \\
 &\leq
 \sum_{\substack{
   \xx_{\I},\ww_{\I},\cc_{\I}:\\
   \xx_{\I}\notin\E(f_{\I},\cc_{\I})\\
   \ww_{\I}\in\fC_{f_{\I}}(\cc_{\I})
 }}
 \mu_{W_{\I}|X_{\I}}
 (\ww_{\I}|\xx_{\I})
 \mu_{X_{\I}}(\xx_{\I})
 \sum_{i=1}^{|\I|}
 \lrbar{
  \frac1
  {\mu_{W_i|X_i}(\fC_{f_i}(\cc_i)|\xx_i)|\im\F_i|}
  -
  1
 }
 \prod_{k=i+1}^{|\I|}
 \frac1
 {\mu_{W_k|X_k}(\fC_{f_k}(\cc_k)|\xx_k)|\im\F_k|}
 \notag
 \\*
 &\leq
 \sum_{i=1}^{|\I|}
 \sum_{\substack{
   \xx_i,\ww_i,\cc_i:\\
   \xx_i\notin\E(f_i,\cc_i)\\
   \ww_i\in\fC_{f_i}(\cc_i)
 }}
 \mu_{W_i|X_i}
 (\ww_i|\xx_i)
 \mu_{X_i}(\xx_i)
 \lrbar{
  \frac1
  {\mu_{W_i|X_i}(\fC_{f_i}(\cc_i)|\xx_i)|\im\F_i|}
  -
  1
 }
 \notag
 \\
 &=
 \sum_{i=1}^{|\I|}
 \sum_{\substack{
   \xx_i,\cc_i:\\
   \xx_i\notin\E(f_i,\cc_i)
 }}
 \mu_{X_i}(\xx_i)
 \lrbar{
  \frac1
  {|\im\F_i|}
  -
  \mu_{W_i|X_i}(\fC_{f_i}(\cc_i)|\xx_i)
 }
 \notag
 \\
 &=
 \sum_{i=1}^{|\I|}
 \sum_{\substack{
   \xx_i,\cc_i
 }}
 \mu_{X_i}(\xx_i)
 \lrbar{
  \frac1
  {|\im\F_i|}
  -
  \mu_{W_i|X_i}(\fC_{f_i}(\cc_i)|\xx_i)
 }
 -
 \sum_{i=1}^{|\I|}
 \sum_{\substack{
   \xx_i,\cc_i:\\
   \xx_i\in\E(f_i,\cc_i)
 }}
 \frac{\mu_{X_i}(\xx_i)}
 {|\im\F_i|},
\end{align}
where
the second inequality comes from Lemma \ref{lem:diff-prod}
in Appendix \ref{sec:proof-lemma},
and the third inequality comes from the fact that
\begin{align}
 &
 \sum_{\substack{
   \ww_{\I\setminus\{i\}},\cc_{\I\setminus\{i\}}:\\
   \xx_{i'}\notin\E(f_{i'},\cc_{i'})\ \text{for all}\ i'\in\I\setminus\{i\}\\
   \ww_{\I\setminus\{i\}}
   \in\fC_{f_{\I\setminus\{i\}}}(\cc_{\I\setminus\{i\}})
 }}
 \lrB{
  \prod_{k\in\I\setminus\{i\}}
  \mu_{W_k|X_k}
  (\ww_k|\xx_k)
 }
 \prod_{k=i+1}^{|\I|}
 \frac1
 {\mu_{W_k|X_k}(\fC_{f_k}(\cc_k)|\xx_k)|\im\F_k|}
 \notag
 \\*
 &\leq
 \lrB{
  \prod_{k=1}^{i-1}
  \sum_{\substack{
    \ww_k,\cc_k:\\
    \ww_k\in\fC_{f_k}(\cc_k)
  }}
  \mu_{W_k|X_k}(\ww_k|\xx_k)
 }
 \lrB{
  \prod_{k=i+1}^{|\I|}
  \sum_{\substack{
    \ww_k,\cc_k:\\
    \xx_k\notin\E(f_k,\cc_k)\\
    \ww_k\in\fC_{f_k}(\cc_k)
  }}
  \frac{\mu_{W_k|X_k}(\ww_k|\xx_k)}
  {\mu_{W_k|X_k}(\fC_{f_k}(\cc_k)|\xx_k)|\im\F_k|}
 }
 \notag
 \\
 &\leq
 \prod_{k=i+1}^{|\I|}
 \sum_{\cc_k}
 \frac1
 {|\im\F_k|}
 \notag
 \\
 &=
 1.
\end{align}

The third term on the righthand side of
(\ref{eq:proof-crng-dlc-error})
is evaluated as
\begin{align}
 &
 E_{C_{\I}}
 \lrB{
  \text{the third term of (\ref{eq:proof-crng-dlc-error})}
 }
 \notag
 \\*
 &\leq
 \sum_{\substack{
   \xx_{\I},\yy,\ww_{\I},\hww_{\I},\zz_{\I},\cc_{\I}:\\
   \ww_{\I}\in\fC_{f_{\I}}(\cc_{\I})\\
   \hww_{\I}\neq\ww_{\I}
 }}
 \mu_{Z_{\J}|W_{\I}Y}(\zz_{\J}|\hww_{\I},\yy)
 \mu_{\hW_{\I}|C_{\I}M_{\I}Y}
 (\hww_{\I}|f_{\I}(\ww_{\I}),g_{\I}(\ww_{\I}),\yy)
 \mu_{W_{\I}X_{\I}Y}(\ww_{\I},\xx_{\I},\yy)
 \notag
 \\
 &=
 \sum_{\substack{
   \yy,\ww_{\I},\hww_{\I}:\\
   \hww_{\I}\neq\ww_{\I}
 }}
 \mu_{\hW_{\I}|C_{\I}M_{\I}Y}
 (\hww_{\I}|f_{\I}(\ww_{\I}),g_{\I}(\ww_{\I}),\yy)
 \mu_{W_{\I}Y}(\ww_{\I},\yy).
 \label{eq:proof-crng-dlc-error-Z0}
\end{align}

The fourth term on the righthand side of
(\ref{eq:proof-crng-dlc-error})
is evaluated as
\begin{align}
 E_{C_{\I}}
 \lrB{
  \text{the fourth term of (\ref{eq:proof-crng-dlc-error})}
 }
 &\leq
 \sum_{\substack{
   \xx_{\I},\yy,\ww_{\I},\zz_{\J},\cc_{\I}:\\
   \ww_{\I}\in\fC_{f_{\I}}(\cc_{\I})\\
   (\xx_{\I},\yy,\zz_{\J})\in\E(D_{\J})
 }}
 \mu_{Z_{\J}|W_{\I}Y}
 (\zz_{\J}|\ww_{\I},\yy)
 \mu_{\hW_{\I}|C_{\I}M_{\I}Y}
 (\ww_{\I}|\cc_{\I},g_{\I}(\ww_{\I}),\yy)
 \mu_{W_{\I}X_{\I}Y}
 (\ww_{\I},\xx_{\I},\yy)
 \notag
 \\
 &\leq
 \sum_{\substack{
   \xx_{\I},\yy,\ww_{\I},\zz_{\J},\cc_{\I}:\\
   \ww_{\I}\in\fC_{f_{\I}}(\cc_{\I})\\
   (\xx_{\I},\yy,\zz_{\J})\in\E(D_{\J})
 }}
 \mu_{Z_{\J}|W_{\I}Y}
 (\zz_{\J}|\ww_{\I},\yy)
 \mu_{W_{\I}X_{\I}Y}
 (\ww_{\I},\xx_{\I},\yy)
 \notag
 \\
 &=
 \sum_{\substack{
   \xx_{\I},\yy,\zz_{\J}:\\
   (\xx_{\I},\yy,\zz_{\J})\in\E(D_{\J})
 }}
 \mu_{X_{\I}YZ_{\J}}
 (\xx_{\I},\yy,\zz_{\J}),
  \label{eq:proof-crng-dlc-error-5}
\end{align}
where the second inequality comes from
$\mu_{\hW_{\I}|C_{\I}M_{\I}Y}(\ww_{\I}|\cc_{\I},g_{\I}(\ww_{\I}),\yy)
\leq 1$.

Finally, from Lemmas \ref{lem:channel-bcp} and \ref{lem:channel-crp},
we have
\begin{align}
 E_{(F,G)_{\I}C_{\I}S}\lrB{
	\Error(F_{\I},G_{\I},C_{\I},S)
 }
 &\leq
 \sum_{\substack{
	 i\in\I:
	 |\im\F_i|>1
 }}
 E_{F_i}\lrB{
	\sum_{\substack{
		\xx_i,\cc_i
	}}
	\mu_{X_i}(\xx_i)
	\lrbar{
	 \frac1
	 {|\im\F_i|}
	 -
	 \mu_{W_i|X_i}(\fC_{F_i}(\cc_i)|\xx_i)
	}
 }
 \notag
 \\*
 &\quad
 +
 E_{(F,G)_{\I}}\lrB{
	\sum_{\substack{
    \ww_{\I},\hww_{\I},\yy:\\
    \hww_{\I}\neq\ww_{\I}
  }}
  \mu_{\hW_{\I}|C_{\I}M_{\I}Y}
  (\hww_{\I}|(F,G)_{\I}(\ww_{\I}),\yy)
  \mu_{W_{\I}Y}(\ww_{\I},\yy)
 }
 \notag
 \\*
 &\quad
 +
 \sum_{\substack{
   \xx_{\I},\yy,\zz_{\J}:\\
	 (\xx_{\I},\yy,\zz_{\J})\in\E(D_{\J})
 }}
 \mu_{X_{\I}YZ_{\J}}
 (\xx_{\I},\yy,\zz_{\J})
 \notag
 \\
 \begin{split}
	&\leq
	\sum_{\substack{
		i\in\I:
		r_i>0
	}}
	\sqrt{
	 \alpha_{F_i}-1
	 +
	 [\beta_{F_i}+1]2^{-n\ugamma(i)}
	}
	+
	2
	\sum_{\substack{
		i\in\I:
		r_i>0
	}}
	\mu_{W_iX_i}(\uT_{W_i|X_i}^{\complement})
	\\*
	&\quad
	+2\sum_{
	 \I'\in2^{\I}\setminus\{\emptyset\}
	}
	\alpha_{(F,G)_{\I'}}\lrB{\beta_{(F,G)_{\Ipc}}+1}
	2^{
	 -n\ogamma(\I')
	}
	+2\beta_{(F,G)_{\I}}
	+2\mu_{W_{\I}Y}(\oT_{W_{\I}|Y}^{\complement})
	\\*
	&\quad
	+
	\sum_{i\in\I}
	\Prob\lrsb{
	 d^{(n)}(X^n_{\I},Y^n, Z^n_j)>D_j+\delta
	}
 \end{split}
 \label{eq:proof-lossy-error}
\end{align}
where $r_i$ and $R_i$ are defined by (\ref{eq:ri}) and (\ref{eq:Ri}),
respectively,
\begin{align*}
 \uT_{W_i|X_i}
 &\equiv
 \lrb{
  (\ww_i,\xx_i):
  \begin{aligned}
   \frac 1n
   \log_2\frac1{
    \mu_{W_i|X_i}(\ww_i|\xx_i)
   }
   \geq
   \uH(\WW_i|\XX_i)-\e
  \end{aligned}
 }
 \\
 \oT_{W_{\I}|Y}
 &\equiv
 \lrb{
  (\ww_{\I},\yy):
  \begin{aligned}
    \frac 1n
   \log_2\frac1{
    \mu_{W_{\I'}|W_{\Ipc}Y}(\ww_{\I'}|\ww_{\Ipc},\yy)
   }
   \leq
   \oH(\WW_{\I'}|\WW_{\Ipc},\YY)+\e
   \ \text{for all}\ \I'\in 2^{\I}\setminus\{\emptyset\}
  \end{aligned}
 }
 \\
 \ugamma(i)
 &\equiv
 \uH(\WW_i|\XX_i)-r_i-\e
 \\
 \ogamma(\I')
 &\equiv
 \sum_{i\in\I'}[r_i+R_i]-\oH(\WW_{\I'}|\WW_{\Ipc},\YY)-\e,
\end{align*}
and the first inequality comes from the fact that
$r_i=0$ implies $|\im\F_i|=1$ and
$\mu_{W_i|X_i}(\fC_{F_i}(\cc_i)|\xx_i)=1$.
From (\ref{eq:crng-dXY>D}),
the last term on the righthand side of (\ref{eq:proof-lossy-error})
goes to zero as $n\to\infty$.
From the definitions of limit inferior/superior in probability,
we have the fact that
$\mu_{X_iW_i}(\uT_{W_i|X_i}^{\complement})\to0$,
$\mu_{W_{\I}Y}(\oT_{W_{\I}|Y}^{\complement})\to0$
as $n\to\infty$.
By assuming that
$\{(\F_{\I,n},p_{F_{\I},n})\}_{n=1}^{\infty}$ and
$\{(\G_{\I,n},p_{G_{\I,n}})\}_{n=1}^{\infty}$
have the hash property (Appendix \ref{sec:hash}),
we have
$\alpha_{F_i}\to1$,
$\beta_{F_i}\to0$,
$\alpha_{(F,G)_{\I'}}\to1$,
$\beta_{(F,G)_{\I'}}\to0$
as $n\to\infty$
for all $i\in\I$ and $\I'\in2^{\I}\setminus\{\emptyset\}$.
Since we can take $\e>0$ to be sufficiently small
so as to satisfy $\ugamma(i)>0$ for all $i\in\I$
and $\ogamma(\I')>0$ for all $\I'\in2^{\I}\setminus\{\emptyset\}$
under the conditions (\ref{eq:CRNG-ri}) and (\ref{eq:CRNG-sum[Ri+ri]}),
we have the fact that for sufficiently large $n$ 
there are functions
$\{f_i\}_{i\in\I}$ and $\{g_i\}_{i\in\I}$,
and vectors $\{\cc_i\}_{i\in\I}$, $\bs$
satisfying $\cc_i\in\im\F_i$
and $\Error(f_{\I},g_{\I},\cc_{\I},\bs)\leq\delta$.
\hfill\IEEEQED

\appendix
\subsection{Information-Spectrum Methods}
\label{sec:ispec}

This section introduces the information spectrum methods
introduced in \cite{HAN,HV93}.
Let $\{U_n\}_{n=1}^{\infty}$ be a general sequence of random variables,
where we do not assume conditions such as stationarity and ergodicity.
The generality of
the discussion does not change regardless of whether we assume or not
that the alphabet of $U_n$ is the Cartesian product,
that is, $\U_n\equiv\U^n$.
We do not assume the consistency of $\{U_n\}_{n=1}^{\infty}$
when $\U_n\equiv\U^n$.

First, we review the definition of the limit superior/inferior in
probability.
For a general source $\{U_n\}_{n=1}^{\infty}$,
the {\em limit superior in probability} $\plimsupn U_n$
and the {\em limit inferior in probability} $\pliminfn U_n$ are defined
as
\begin{align*}
 \plimsupn U_n
 &\equiv 
 \inf\lrb{\theta: \limn \Prob\lrsb{U_n>\theta}=0}
 \\
 \pliminfn U_n
 &\equiv
 \sup\lrb{\theta: \limn \Prob\lrsb{U_n<\theta}=0}.
\end{align*}

In the following, we introduce the key inequalities used in the
proof of the converse part.
We have the following relations~\cite[Section 1.3]{HAN}:
\begin{align}
 \plimsupn U_n
 &\geq\pliminfn U_n
 \label{eq:plimsup>pliminf}
 \\
 \plimsupn\lrB{U_n+V_n}
 &\leq \plimsupn U_n + \plimsupn V_n
 \label{eq:plimsup-upper}
 \\
 \plimsupn\lrB{U_n+V_n}
 &\geq \plimsupn U_n + \pliminfn V_n
 \label{eq:plimsup-lower}
 \\
 \pliminfn\lrB{U_n+V_n}
 &\leq \plimsupn U_n + \pliminfn V_n
 \label{eq:pliminf-upper}
 \\
 \pliminfn\lrB{U_n+V_n}
 &\geq \pliminfn U_n + \pliminfn V_n
 \label{eq:pliminf-lower}
 \\
 \plimsupn U_n
 &=
 -\pliminfn[-U_n].
 \label{eq:plimsup-pliminf}
\end{align}
Here, we show the lemma for a sequence of constant random variables.
\begin{lem}
\label{lem:constant}
For a sequence of constant random variables
$\{U_n\}_{n=1}^{\infty}=\{u_n\}_{n=1}^{\infty}$,
we have
\begin{align*}
 \plimsupn U_n
 &= 
 \limsupn u_n.
\end{align*}
\end{lem}
\begin{IEEEproof}
We have
\begin{align}
 \plimsupn U_n
 &= 
 \inf\lrb{\theta: \limn \Prob\lrsb{U_n>\theta}=0}
 \notag
 \\
 &=
 \inf\lrb{\theta:
	\begin{aligned}
	 &\Prob\lrsb{U_n>\theta}<\e
	 \ \text{for all $\e>0$}\\
	 &\text{and all sufficiently large $n$}
	\end{aligned}
 }
 \notag
 \\
 &=
 \inf\lrb{\theta:
	\begin{aligned}
	 &\Prob\lrsb{U_n>\theta}=0\\
	 &\text{for all sufficiently large $n$}
	\end{aligned}
 }
 \notag
 \\
 &=
 \inf\lrb{\theta:
	u_n\leq \theta\ 
	\text{for all sufficiently large $n$}
 }
 \notag
 \\
 &=
 \limsupn u_n,
 \end{align}
 where the third and the fourth equality come from the fact that
 $U_n=u_n$ with probability $1$
 and from this we have $\Prob(U_n>\theta)\in\{0,1\}$.
\end{IEEEproof}

Next, let $\UU\equiv\{U_n\}_{n=1}^{\infty}$
be a general sequence of random variables,
which is called a {\em general source}.
For sequence $\{\mu_{U_n}\}_{n=1}^{\infty}$
of probability distributions corresponding to $\UU$,
we define the {\em spectral sup-entropy rate} $\oH(\UU)$ as
\begin{align*}
 \oH(\UU)
 &\equiv
 \plimsupn\frac 1n\log_2\frac1{\mu_{U_n}(U_n)}.
\end{align*}
For general sequence $\{\mu_{U_nV_n}\}_{n=1}^{\infty}$ of
joint probability distributions
corresponding to $(\UU,\VV)=\{(U_n,V_n)\}_{n=1}^{\infty}$,
we define the {\em spectral conditional sup-entropy rate} $\oH(\UU|\VV)$,
the {\em spectral conditional inf-entropy rate} $\uH(\UU|\VV)$,
the {\em spectral sup-information rate} $\oI(\UU;\VV)$,
and the {\em spectral inf-information rate} $\uI(\UU;\VV)$
as
\begin{align*}
 \oH(\UU|\VV)
 &\equiv
 \plimsupn\frac 1n\log_2\frac1{\mu_{U_n|V_n}(U_n|V_n)}
 \\
 \uH(\UU|\VV)
 &\equiv
 \pliminfn\frac 1n\log_2\frac1{\mu_{U_n|V_n}(U_n|V_n)}
 \\
 \oI(\UU;\VV)
 &\equiv
 \plimsupn\frac 1n\log_2\frac{\mu_{U_n|V_n}(U_n|V_n)}{\mu_{U_n}(U_n)}
 \\
 \uI(\UU;\VV)
 &\equiv
 \pliminfn\frac 1n\log_2\frac{\mu_{U_n|V_n}(U_n|V_n)}{\mu_{U_n}(U_n)}.
\end{align*}

The following lemma is related to the non-negativity of the divergence between two distributions.
\begin{lem}[{\cite[Lemma 3.2.1, Definition 4.1.3]{HAN}}]
\label{lem:pliminf-div}
Let $\{\mu_{U_n}\}_{n=1}^{\infty}$
be a general sequence of probability distributions
corresponding to $\UU\equiv\{U_n\}_{n=1}^{\infty}$.
For each $n$,
let $\nu_n$ be an arbitrary probability distribution on $\U_n$.
Then we have
\begin{equation*}
 \pliminfn\frac 1n\log_2\frac{\mu_{U_n}(U_n)}{\nu_n(U_n)}\geq 0.
\end{equation*}
\end{lem}
\begin{IEEEproof}
For completeness, this paper proves the lemma
by following the proof given by \cite[Lemma 3.2.1, Definition 4.1.3]{HAN}.

For a given $\gamma>0$, we define $\E$ as
\begin{equation*}
 \E
 \equiv
 \lrb{
  \uu:
  \frac1n\log_2\frac{\mu_{U_n}(\uu)}{\nu_n(\uu)}
  \leq -\gamma
 }.
\end{equation*}
Then we have
\begin{align}
 \Prob\lrsb{
  \frac1n\log_2\frac{\mu_{U_n}(U_n)}{\nu_n(U_n)}
  \leq -\gamma
 }
 &=
 \sum_{\uu\in\E}
 \mu_{U_n}(\uu)
 \notag
 \\
 &\leq
 \sum_{\uu\in\E}
 \nu_n(\uu)2^{-n\gamma}
 \notag
 \\
 &\leq
 2^{-n\gamma}
\end{align}
which implies
\begin{equation*}
 \limn\Prob\lrsb{
  \frac1n\log_2\frac{\mu_{U_n}(U_n)}{\nu_n(U_n)}
  \leq -\gamma
 }
 =0.
\end{equation*}
From this inequality and the definition of $\pliminfn$, we have
\begin{equation*}
 \pliminfn\frac 1n\log_2\frac{\mu_{U_n}(U_n)}{\nu_n(U_n)}
 \geq -\gamma.
\end{equation*}
Then the lemma is verified by letting $\gamma\to0$.
\end{IEEEproof}

We show the following lemmas, which are used in the proof of
the converse theorem.
\begin{lem}
\label{lem:oH>uH>0}
\begin{equation*}
 \oH(\UU|\VV)\geq\uH(\UU|\VV)\geq 0.
 \label{eq:oH>uH>0}
\end{equation*}
\end{lem}
\begin{IEEEproof}
The first inequality comes from (\ref{eq:plimsup>pliminf}).
The second inequality comes from the fact that
$\mu_{U_n|V_n}(U_n|V_n)\leq 1$.
\end{IEEEproof}

\begin{lem}[{\cite[Lemma1]{CRNG-MULTI}}]
\label{lem:bound-by-cardinality}
Let $\U_n$ be the alphabet of $U^n$. Then
\begin{equation*}
 \oH(\UU)
 \leq
 \limsupn\frac{\log_2|\U_n|}n.
\end{equation*}
\end{lem}
\begin{IEEEproof}
Let $\nu_n$ be a uniform distribution on $\U_n$.
Then we have
\begin{align}
 \limsupn\frac{\log_2|\U_n|}n-\oH(\UU)
 &=
 \plimsupn\frac1n\log_2\frac1{\nu_n(U_n)}
 -\plimsupn\frac1n\log_2\frac1{\mu_{U_n}(U_n)}
 \notag
 \\
 &=
 \plimsupn\frac1n\log_2\frac1{\nu_n(U_n)}
 +\pliminfn\frac1n\log_2\mu_{U_n}(U_n)
 \notag
 \\
 &\geq
 \pliminfn\frac 1n\log_2\frac{\mu_{U_n}(U_n)}{\nu_n(U_n)}
 \notag
 \\
 &\geq
 0,
\end{align}
where the first equality comes from Lemma \ref{lem:constant}
and the fact that 
$\frac1n\log_2(1/\nu_n(U_n))$ is a constant random variable
satisfying $\frac1n\log_2(1/\nu_n(U_n))=\frac1n\log_2|\U_n|$,
the second equality comes from (\ref{eq:plimsup-pliminf}),
the first inequality comes from (\ref{eq:pliminf-upper}),
and the second inequality comes from  Lemma~\ref{lem:pliminf-div}.
\end{IEEEproof}

\begin{lem}
\label{lem:oH(UU'|V)<oH(U'|UV)+oH(U|V)}
For a triplet of general sources,
$(\UU,\UU',\VV)=\{(U_n,U'_n,V_n)\}_{n=1}^{\infty}$,
we have
\begin{equation*}
 \oH(\UU,\UU'|\VV)\leq \oH(\UU'|\UU,\VV) + \oH(\UU|\VV).
\end{equation*}
\end{lem}
\begin{IEEEproof}
We have
\begin{align}
 \oH(\UU,\UU'|\VV)
 &=
 \plimsupn\frac 1n\log_2
 \frac1{\mu_{U_nU'_n|V_n}(U_n,U'_n|V_n)}
 \notag
 \\
 &=
 \plimsupn\frac 1n\log_2
 \frac
 {\mu_{V_n}(V_n)}
 {\mu_{U_nU'_nV_n}(U_n,U'_n,V_n)}
 \notag
 \\
 &=
 \plimsupn\frac 1n\log_2
 \frac
 {\mu_{V_n}(V_n)}
 {\mu_{U'_n|U_nV_n}(U'_n|U_n,V_n)\mu_{U_nV_n}(U_n,V_n)}
 \notag
 \\
 &=
 \plimsupn\frac 1n\log_2
 \frac
 1
 {\mu_{U'_n|U_nV_n}(U'_n|U_n,V_n)\mu_{U_n|V_n}(U_n|V_n)}
 \notag
 \\
 &\leq
 \plimsupn\frac 1n\log_2
 \frac
 1
 {\mu_{U'_n|U_nV_n}(U'_n|U_n,V_n)}
 +
 \plimsupn\frac 1n\log_2
 \frac
 1
 {\mu_{U_n|V_n}(U_n|V_n)}
 \notag
 \\
 &=
 \oH(\UU'|\UU,\VV)
 +
 \oH(\UU|\VV),
\end{align}
where the inequality comes from (\ref{eq:plimsup-upper}).
\end{IEEEproof}

\begin{lem}
\label{lem:oH(UU'|V)>oH(U|V)}
For a triplet of general sources,
$(\UU,\UU',\VV)=\{(U_n,U'_n,V_n)\}_{n=1}^{\infty}$,
we have
\begin{equation*}
 \oH(\UU,\UU'|\VV)\geq \oH(\UU|\VV).
\end{equation*}
\end{lem}
\begin{IEEEproof}
We have
\begin{align}
 \oH(\UU,\UU'|\VV)
 &=
 \plimsupn\frac 1n\log_2
 \frac1{\mu_{U_nU'_n|V_n}(U_n,U'_n|V_n)}
 \notag
 \\
 &=
 \plimsupn\frac 1n\log_2
 \frac
 {\mu_{V_n}(V_n)}
 {\mu_{U_nU'_nV_n}(U_n,U'_n,V_n)}
 \notag
 \\
 &=
 \plimsupn\frac 1n\log_2
 \frac
 {\mu_{V_n}(V_n)}
 {\mu_{U'_n|U_nV_n}(U'_n|U_n,V_n)\mu_{U_nV_n}(U_n,V_n)}
 \notag
 \\
 &=
 \plimsupn\frac 1n\log_2
 \frac
 1
 {\mu_{U'_n|U_nV_n}(U'_n|U_n,V_n)\mu_{U_n|V_n}(U_n|V_n)}
 \notag
 \\
 &\geq
 \pliminfn\frac 1n\log_2
 \frac
 1
 {\mu_{U'_n|U_nV_n}(U'_n|U_n,V_n)}
 +
 \plimsupn\frac 1n\log_2
 \frac
 1
 {\mu_{U_n|V_n}(U_n|V_n)}
 \notag
 \\
 &=
 \uH(\UU'|\UU,\VV)
 +
 \oH(\UU|\VV)
 \notag
 \\
 &\geq
 \oH(\UU|\VV),
\end{align}
where the first inequality comes from (\ref{eq:plimsup-lower})
and the second inequality comes from
Lemma \ref{lem:oH>uH>0}.
\end{IEEEproof}

\begin{lem}[{\cite[Lemma2]{CRNG-MULTI}}]
\label{lem:oH(U|V)>oH(U|VV')}
For a triplet of general sources,
$(\UU,\VV,\VV')=\{(U_n,V_n,V'_n)\}_{n=1}^{\infty}$,
we have
\begin{equation*}
 \oH(\UU|\VV)\geq \oH(\UU|\VV,\VV').
\end{equation*}
\end{lem}
\begin{IEEEproof}
We have
\begin{align}
 \oH(\UU|\VV)-\oH(\UU|\VV,\VV')
 &=
 \plimsupn\frac 1n\log_2
 \frac1{\mu_{U_n|V_n}(U_n|V_n)}
 -
 \plimsupn\frac 1n\log_2\frac1{\mu_{U_n|V_nV'_n}(U_n|V_n,V'_n)}
 \notag
 \\
 &=
 \plimsupn\frac 1n\log_2
 \frac1{\mu_{U_n|V_n}(U_n|V_n)}
 +
 \pliminfn\frac 1n\log_2\mu_{U_n|V_nV'_n}(U_n|V_n,V'_n)
 \notag
 \\
 &\geq
 \pliminfn\frac 1n\log_2
 \frac{\mu_{U_n|V_nV'_n}(U_n|V_n,V'_n)}{\mu_{U_n|V_n}(U_n|V_n)}
 \notag
 \\
 &=
 \pliminfn\frac 1n\log_2
 \frac{\mu_{U_nV_nV'_n}(U_n,V_n,V'_n)}
 {\mu_{U_n|V_n}(U_n|V_n)\mu_{V_nV'_n}(V_n,V'_n)}
 \notag
 \\
 &\geq
 0,
\end{align}
where the second equality comes from (\ref{eq:plimsup-pliminf}),
the first inequality comes from (\ref{eq:pliminf-upper}),
and the second inequality comes from Lemma~\ref{lem:pliminf-div}.
\end{IEEEproof}

The following lemma
means that
the information\footnote{In fact, $\oI(\VV;\VV|\VV')=\oH(\VV|\VV')$.}
of $\VV$ for given $\VV'$
is greater than the uncertainty of $\UU$ for given $\VV'$
before the observation of $\VV$,
when complimentary information $\VV$ eliminates
the uncertainty of $\UU$ for given $\VV'$.
\begin{lem}
\label{lem:sw-bound}
For the triplet of general sources $(\UU,\VV,\VV')=\{(U_n,V_n,V'_n)\}_{n=1}^{\infty}$
satisfying
$\oH(\UU|\VV,\VV')=0$,
we have
\begin{equation*}
 \oH(\VV|\VV')\geq\oH(\UU|\VV').
\end{equation*}
\end{lem}
\begin{IEEEproof}
We have
\begin{align}
 \oH(\VV|\VV')
 &=
 \plimsupn\frac1n\log\frac1{\mu_{V_nV'_n}(V_n|V'_n)}
 \notag
 \\
 &=
 \plimsupn
 \frac1n\log
 \frac{
	\mu_{V'_n}(V'_n)
	\mu_{U_nV_nV'_n}(U_n,V_n,V'_n)
 }{
	\mu_{V_nV'_n}(V_n,V'_n)\mu_{U_nV_nV'_n}(U_n,V_n,V'_n)
 }
 \notag
 \\
 &=
 \plimsupn\frac1n
 \log\frac{
	\mu_{U_n|V_nV'_n}(U_n|V_n,V'_n)
 }{
	\mu_{V_n|U_nV'_n}(V_n|U_n,V'_n)
	\mu_{U_n|V'_n}(U_n|V'_n)
 }
 \notag
 \\
 &\geq
 \plimsupn\frac1n
 \log\frac{
	1
 }{
	\mu_{U_n|V'_n}(U_n|V'_n)
 }
 +
 \pliminfn\frac1n
 \log\frac{
	\mu_{U_n|V_nV'_n}(U_n|V_n,V'_n)
 }{
	\mu_{V_n|U_nV'_n}(V_n|U_n,V'_n)
 }
 \notag
 \\
 &\geq
 \plimsupn\frac1n
 \log\frac{
	1
 }{
	\mu_{U_n|V'_n}(U_n|V'_n)
 }
 +
 \pliminfn\frac1n
 \log\frac{
	1
 }{
	\mu_{V_n|U_nV'_n}(V_n|U_n,V'_n)
 }
 +
 \pliminfn\frac1n
 \log \mu_{U_n|V_nV'_n}(U_n|V_n,V'_n)
 \notag
 \\
 &=
 \plimsupn\frac1n
 \log\frac{
	1
 }{
	\mu_{U_n|V'_n}(U_n|V'_n)
 }
 +
 \pliminfn\frac1n
 \log\frac{
	1
 }{
	\mu_{V_n|U_nV'_n}(V_n|U_n,V'_n)
 }
 -
 \plimsupn
 \frac1n
 \log\frac{
	1
 }{
	\mu_{U_n|V_nV'_n}(U_n|V_n,V'_n)
 }
 \notag
 \\
 &=
 \oH(\UU|\VV')
 +\uH(\VV|\UU,\VV')
 -\oH(\UU|\VV,\VV')
 \notag
 \\
 &\geq
 \oH(\UU|\VV'),
\end{align}
where the first inequality comes from (\ref{eq:plimsup-lower}),
the second inequality comes from (\ref{eq:pliminf-lower}),
the next equality comes from (\ref{eq:plimsup-pliminf}),
and the last inequality comes from Lemma \ref{lem:oH>uH>0}
and the assumption $\oH(\UU|\VV,\VV')=0$.
\end{IEEEproof}

We introduce the following lemma, which is analogous to Fano inequality.
\begin{lem}[{\cite[Lemma 4]{K08}\cite[Lemma 7]{CRNG}}]
\label{lem:fano}
Let $(\UU,\VV)\equiv\{(U_n,V_n)\}_{n=1}^{\infty}$
be a sequence of two random variables.
If there is a sequence $\{\psi_n\}_{n=1}^{\infty}$ of functions
satisfying the condition
\begin{equation}
 \limn \Prob(\psi_n(V_n)\neq U_n)=0,
 \label{eq:fano-error}
\end{equation}
then
\begin{equation*}
 \oH(\UU|\VV)=0.
\end{equation*}
\end{lem}
\begin{IEEEproof}
We introduce this lemma following the proof of \cite[Lemma 1.3.2]{HAN} for the completeness of this paper.

Let $\{\psi_n\}_{n=1}^{\infty}$ be a sequence of deterministic
functions satisfying
(\ref{eq:fano-error}).
For $\gamma>0$, let
\begin{align*}
 \E
 &\equiv\lrb{
	(\uu,\vv): \psi_n(\vv)\neq\uu
 }
 \\
 \E'
 &\equiv\lrb{
  (\uu,\vv): \frac 1n\log_2\frac1{\mu_{U_n|V_n}(\uu|\vv)}\geq\gamma
 }.
\end{align*}
Then we have
\begin{align}
 \Prob\lrsb{
	\frac 1n\log_2\frac1{\mu_{U_n|V_n}(U_n|V_n)} > \gamma
 }
 &\leq
 \mu_{U_nV_n}(\E')
 \notag
 \\
 &=
 \mu_{U_nV_n}(\E\cap\E')
 +\mu_{U_nV_n}(\E^{\complement}\cap\E')
 \notag
 \\
 &=
 \mu_{U_nV_n}(\E\cap\E')
 +\sum_{(\uu,\vv)\in\E^{\complement}\cap\E'}\mu_{U_nV_n}(\uu,\vv)
 \notag
 \\
 &=
 \mu_{U_nV_n}(\E\cap\E')
 +
 \sum_{\vv\in\V_n}\mu_{V_n}(\vv)
 \sum_{\substack{
	 \uu\in\U_n:\\
	 \psi_n(\vv)=\uu\\
	 (\uu,\vv)\in\E'
 }}
 \mu_{U_n|V_n}(\uu|\vv)
 \notag
 \\
 &\leq
 \mu_{U_nV_n}(\E)
 +
 \sum_{\vv\in\V_n}\mu_{V_n}(\vv)
 \sum_{\uu\in\U_n: \psi_n(\vv)=\uu}
 2^{-n\gamma}
 \notag
 \\
 &=
 \Prob(\psi_n(V_n)\neq U_n)+2^{-n\gamma},
\end{align}
where the second inequality comes from the definition of $\E'$
and the last equality comes from the fact that
for all $\vv$ there is a unique $\uu$ satisfying $\psi_n(\vv)=\uu$.
From this inequality and (\ref{eq:fano-error}), we have
\begin{equation*}
 \limn\Prob\lrsb{
	\frac 1n\log_2\frac1{\mu_{U_n|V_n}(U_n|V_n)}>\gamma
 }=0.
\end{equation*}
Then we have
\begin{equation*}
 0\leq\oH(\UU|\VV)\leq \gamma
\end{equation*}
from the definition of $\oH(\UU|\VV)$.
We have $\oH(\UU|\VV)=0$ by letting $\gamma\to0$.
\end{IEEEproof}

\subsection{Expectation of Constrained-Random Number Generator}
\label{sec:crng-expectation}

The following fact is used in the proof of Theorem \ref{thm:crng}.
\begin{lem}
Let us assume that 
the constrained-random number generator
(deterministic function) $\mathrm{crng}:\cS\to\U$
generates random number $U=\mathrm{crng}(S)$
by using random source $S$.
Then we have
\begin{equation*}
 E_S\lrB{
  \lambda(\mathrm{crng}(S))
 }
 =
 E_U\lrB{\lambda(U)}
\end{equation*}
for any (integrable) function $\lambda$ on $\U$.
\end{lem}
\begin{IEEEproof}
We have
\begin{align}
 E_S\lrB{
  \lambda(\mathrm{crng}(S))
 }
 &=
 \sum_{s\in\cS}
 \mu_S(s)
 \lambda(\mathrm{crng}(s))
 \notag
 \\
 &=
 \sum_{s\in\cS}
 \mu_S(s)
 \sum_{u\in\U}
 \lambda(u)
 \chi(\mathrm{crng}(s)=u)
 \notag
 \\
 &=
 \sum_{u\in\U}
 \lambda(u)
 \sum_{s\in\cS}
 \mu_S(s)
 \chi(\mathrm{crng}(s)=u)
 \notag
 \\
 &=
 \sum_{u\in\U}
 \lambda(u)
 \mu_U(u)
 \notag
 \\
 &=
 E_U\lrB{\lambda(U)}
\end{align}
for any (integrable) function $\lambda$ on $\U$.
\end{IEEEproof}

\subsection{$(\aalpha,\bbeta)$-hash property}
\label{sec:hash}

In this section, we review the hash property
introduced in \cite{CRNG}\cite{HASH-BC} and show two basic lemmas.
For set $\F$ of functions,
let $\im\F\equiv \bigcup_{f\in\F}\{f(\ww): \ww\in\W^n\}$.

\begin{df}[{\cite[Definition~3]{CRNG}}]
Let $\F_n$ be a set of functions on $\U^n$.
For probability distribution $p_{F_n}$ on $\F_n$, we
call the pair $(\F_n,p_{F_n})$ an {\em ensemble}.
Then, $(\F_n,p_{F_n})$ has
$(\alpha_{F_n},\beta_{F_n})$-{\em hash property} if
there is a pair $(\alpha_{F_n},\beta_{F_n})$
depending on $p_{F_n}$ such that
\begin{align}
 \sum_{\substack{
	 \ww'\in\W^n\setminus\{\ww\}:
	 \\
	 p_{F_n}(\{f: f(\ww) = f(\ww')\})>\frac{\alpha_{F_n}}{|\im\F_n|}
 }}
 p_{F_n}\lrsb{\lrb{f: f(\ww) = f(\ww')}}
 \leq
 \beta_{F_n}
 \label{eq:hash}
\end{align}
for any $\ww\in\W^n$.
Consider the following conditions for two sequences
$\aalpha_F\equiv\{\alpha_{F_n}\}_{n=1}^{\infty}$ and
$\bbeta_F\equiv\{\beta_{F_n}\}_{n=1}^{\infty}$,
\begin{align}
 \limn \alpha_{F_n}
 &=1
 \label{eq:alpha}
 \\
 \limn \beta_{F_n}
 &=0.
 \label{eq:beta}
\end{align}
Then, we say that 
$(\bcF,\bp_F)$ has $(\aalpha_F,\bbeta_F)$-{\em hash property}
if $\aalpha_F$ and $\bbeta_F$
satisfy (\ref{eq:hash})--(\ref{eq:beta}).
Throughout this paper,
we omit the dependence of $\F$ and $F$ on $n$.
\end{df}

It should be noted that
when $\F$ is a $2$-universal class of hash functions \cite{CW79}
and  $p_F$ is the uniform distribution on $\F$,
then $(\bcF,\bp_F)$ has $(\one,\zero)$-hash property.
Random binning \cite{C75}
and the set of all linear functions \cite{CSI82} are
$2$-universal classes of hash functions.
It is proved in \cite[Section III-B]{HASH-BC} that
an ensemble of sparse matrices (with logarithmic column degree)
has hash property.

First, we introduce the lemma for a joint ensemble.
\begin{lem}[
 {\cite[Lemma 4 of the extended version]{HASH-BC}\cite[Lemma 3]{CRNG}}
]
\label{lem:hash-FG}
Let $(\F,p_F)$ and $(\G,p_G)$ be ensembles of functions
on the same set $\W^n$.
Assume that $(\F,p_F)$ (resp. $(\G,p_G)$) has an $(\alpha_F,\beta_F)$-hash
(resp. $(\alpha_G,\beta_G)$-hash) property.
Let $(f,g)\in\F\times\G$ be a function defined as
\begin{equation*}
 (f,g)(\ww)\equiv(f(\ww),g(\ww))\quad\text{for each}\ \ww\in\W^n.
\end{equation*}
Let $p_{(F,G)}$  be a joint distribution on $\F\times\G$ defined as
\begin{equation*}
 p_{(F,G)}(f,g)\equiv p_F(f)p_G(g)\quad\text{for each}\ (f,g)\in\F\times\G.
\end{equation*}
Then ensemble $(\F\times\G, p_{(F,G)})$ has 
$(\alpha_{(F,G)},\beta_{(F,G)})$-hash property,
where $\alpha_{(F,G)}$ and $\beta_{(F,G)}$ are defined as
\begin{align*}
 \alpha_{(F,G)}
 &\equiv
 \alpha_F\alpha_G
 \\
 \beta_{(F,G)}
 &\equiv
 \beta_F+\beta_G.
\end{align*}
\end{lem}
\begin{IEEEproof}
We show this lemma for the completeness of this paper.
Let
\begin{align*}
 p_{F,\ww,\ww'}&\equiv p_F(\{f: f(\ww)=f(\ww')\})
 \\
 p_{G,\ww,\ww'}&\equiv p_G(\{g: g(\ww)=g(\ww')\})
 \\
 p_{(F,G),\ww,\ww'}&\equiv p_{(F,G)}(
  \{(f,g): (f,g)(\ww)=(f,g)(\ww')\}
 ).
\end{align*}
Then we have
\begin{align}
 &
 \sum_{\substack{
	 \ww'\in\W^n\setminus\{\ww\}:
	 \\
	 p_{(F,G),\ww,\ww'}>\frac{\alpha_{(F,G)}}{|\im\F\times\G|}
 }}
 p_{(F,G)}(\{(f,g): (f,g)(\ww)=(f,g)(\ww')\})
 \notag
 \\*
 &\leq
 \sum_{\substack{
   \ww'\in\W^n\setminus\{\ww\}:
   \\
   p_{F,\ww,\ww'}p_{G,\ww,\ww'}>\frac{\alpha_F\alpha_G}{|\im\F||\im\G|}
 }}
 p_{F,\ww,\ww'}p_{G,\ww,\ww'}
 \notag
 \\
 &=
 \sum_{\substack{
   \ww'\in\W^n\setminus\{\ww\}:
   \\
   p_{F,\ww,\ww'}p_{G,\ww,\ww'}>\frac{\alpha_F\alpha_G}{|\im\F||\im\G|}
   \\
   p_{F,\ww,\ww'}>\frac{\alpha_F}{|\im\F|}
 }}
 p_{F,\ww,\ww'}p_{G,\ww,\ww'}
 +
 \sum_{\substack{
   \ww'\in\W^n\setminus\{\ww\}:
   \\
   p_{F,\ww,\ww'}p_{G,\ww,\ww'}>\frac{\alpha_F\alpha_G}{|\im\F||\im\G|}
   \\
   p_{F,\ww,\ww'}\leq\frac{\alpha_F}{|\im\F|}
 }}
 p_{F,\ww,\ww'}p_{G,\ww,\ww'}
 \notag
 \\
 &\leq
 \sum_{\substack{
   \ww'\in\W^n\setminus\{\ww\}:
   \\
   p_{F,\ww,\ww'}>\frac{\alpha_F}{|\im\F|}
 }}
 p_{F,\ww,\ww'}p_{G,\ww,\ww'}
 +
 \sum_{\substack{
   \ww'\in\W^n\setminus\{\ww\}:
   \\
   p_{G,\ww,\ww'}>\frac{\alpha_G}{|\im\G|}
 }}
 p_{F,\ww,\ww'}p_{G,\ww,\ww'}
 \notag
 \\
 &\leq
 \sum_{\substack{
   \ww'\in\W^n\setminus\{\ww\}:
   \\
   p_{F,\ww,\ww'}>\frac{\alpha_F}{|\im\F|}
 }}
 p_{F,\ww,\ww'}
 +
 \sum_{\substack{
   \ww'\in\W^n\setminus\{\ww\}:
   \\
   p_{G,\ww,\ww'}>\frac{\alpha_G}{|\im\G|}
 }}
 p_{G,\ww,\ww}
 \notag
 \\
 &=
 \beta_F+\beta_G
 \notag
 \\
 &=
 \beta_{(F,G)},
\end{align}
where the first inequality comes from the fact that
$F$ and $G$ are mutually independent
and $\im\F\times\G\subset\im\F\times\im\G$,
and the last inequality comes from the fact that
$p_{F,\ww,\ww'}\leq 1$ and $p_{G,\ww,\ww'}\leq 1$.
Then we have the fact that
$(\F\times\G,p_{(F,G)})$ has 
$(\alpha_{(F,G)},\beta_{(F,G)})$-hash property.
\end{IEEEproof}

The following lemma is related to the {\em balanced-coloring property},
which is analogous to the leftover hash lemma~\cite{IZ89}
and the balanced-coloring
lemma~\cite[Lemma 3.1]{AC98}\cite[Lemma 17.3]{CK11}.
This lemma implies that there is an assignment that divides a set equally.
\begin{lem}[{\cite[Lemma 4]{CRNG}\cite[Lemma 4]{HASH-WTC}}]
\label{lem:mBCP}
Let $\F$ be a set of functions on $\W^n$
and $p_F$ be the probability distribution on $\F$,
where $(\F,p_F)$ satisfies (\ref{eq:hash}).
Then
\begin{align*}
 E_F\lrB{
	\sum_{\cc}
	\lrbar{
	 \frac{Q(\T\cap\fC_F(\cc))}
   {Q(\T)}
   -
   \frac1
   {|\im\F|}
  }
 }
 &\leq
 \sqrt{
  \alpha_F-1
  +
  \frac{
   \lrB{\beta_F+1}|\im\F|
   \max_{\ww\in\T}Q(\ww)
  }
  {Q(\T)}
 }
\end{align*}
for any function $Q:\W^n\to[0,\infty)$ and $\T\subset\W^n$.
\end{lem}
\begin{IEEEproof}
We show this lemma for the completeness of the paper.

First, let
$p_{\ww,\ww'}
\equiv
p_F\lrsb{\lrb{
	f:
	f(\ww)=f(\ww')
}}$.
Then we have
\begin{align}
 \sum_{\substack{
	 \ww\in\T:
	 \\
   p_{\ww,\ww'}
   >\frac{\alpha_F}{|\im\F|}
 }}
 Q(\ww)
 p_{\ww,\ww'}
 &\leq
 \lrB{
  \sum_{\substack{
    \ww\in\W^n:
    \\
    p_{\ww,\ww}
    >\frac{\alpha_F}{|\im\F|}
  }}
  p_{\ww,\ww'}
 }
 \max_{\ww\in\T}Q(\ww)
 \notag
 \\
 &=
 \lrB{
  \sum_{\substack{
    \ww\in\W^n\setminus\{\ww'\}:
    \\
    p_{\ww,\ww'}
    >\frac{\alpha_F}{|\im\F|}
  }}
  p_{\ww,\ww'}
  +
  p_{\ww',\ww'}
 }
 \max_{\ww\in\T}Q(\ww)
 \notag
 \\
 &\leq
 \lrB{\beta_F+1}
 \max_{\ww\in\T}Q(\ww)
 \label{eq:lemma-multi-subset}
\end{align}
for all $\ww'\in\T$,
where the second inequality comes from 
(\ref{eq:hash}) and the fact that $p_{\ww',\ww'}=1$.
Then we have
\begin{align}
 \sum_{\ww\in\T}Q(\ww)
 p_{\ww,\ww'}
 &=
 \sum_{\substack{
	 \ww\in\T:
	 \\
   p_{\ww,\ww'}
   \leq\frac{\alpha_F}{|\im\F|}
 }}
 Q(\ww)
 p_{\ww_s,\ww'_s}
 +
 \sum_{\substack{
   \ww\in\T:
   \\
   p_{\ww,\ww'}
   >\frac{\alpha_F}{|\im\F|}
 }}
 Q(\ww)
 p_{\ww_s,\ww'_s}
 \notag
 \\
 &\leq
 \sum_{\substack{
   \ww\in\T:
   \\
   p_{\ww,\ww'}
   \leq\frac{\alpha_F}{|\im\F|}
 }}
 \frac{\alpha_FQ(\ww)}
 {|\im\F|}
 +
 \lrB{\beta_F+1}
 \max_{\ww\in\T}Q(\ww)
 \notag
 \\
 &=
 \frac{\alpha_FQ(\T)}
 {|\im\F|}
 +
 \lrB{\beta_F+1}
 \max_{\ww\in\T}Q(\ww).
 \label{eq:proof-BCP-multi2}
\end{align}

Next, let $C$ be the random variable subject to the uniform
distribution on $\im\F$.
Then we have
\begin{align}
 E_{FC}\lrB{
  \lrB{
   \sum_{\ww\in\T}Q(\ww)\chi(F(\ww)=C)
  }^2
 }
 &=
 \sum_{\ww'\in\T}Q(\ww')
 \sum_{\ww\in\T}Q(\ww)
 E_{F_{\cS}C}\lrB{
  \chi(F(\ww)=C)
  \chi(F(\ww')=C)
 }
 \notag
 \\
 &=
 \sum_{\ww'\in\T}Q(\ww')
 \sum_{\ww\in\T}Q(\ww)
 E_F\lrB{
  \chi(F(\ww)=F(\ww'))
  E_C\lrB{\chi(F(\ww)=C)}
 }
 \notag
 \\
 &=
 \frac1{|\im\F|}
 \sum_{\ww'\in\T}Q(\ww')
 \sum_{\ww\in\T}Q(\ww)
 p_{\ww,\ww'}
 \notag
 \\
 &\leq
 \frac1{|\im\F|}
 \sum_{\ww'\in\T}Q(\ww')
 \lrB{
  \frac{\alpha_FQ(\T)}
  {|\im\F|}
  +
  [\beta_F+1]\max_{\ww\in\T}Q(\ww)
 }
 \notag
 \\
 &=
 \frac{\alpha_FQ(\T)^2}
 {|\im\F|^2}
 +
 \frac{Q(\T)[\beta_F+1]\max_{\ww\in\T}Q(\ww)}
 {|\im\F|},
 \label{eq:lemma-multi}
\end{align}
where the inequality comes from (\ref{eq:proof-BCP-multi2}).
Then we have
\begin{align}
 &
 E_{FC}\lrB{
	\lrB{
	 \frac{
		Q\lrsb{\T\cap\fC_F(C)}
		|\im\F|
	 }{Q(\T)}
	 -1
	}^2
 }
 \notag
 \\*
 &=
 E_{FC}\lrB{
  \lrB{\sum_{\ww\in\T}
	 \frac{Q(\ww)\chi(F(\ww)=C)
    |\im\F|
   }{Q(\T)}
  }^2
 }
 -2
 E_{FC}\lrB{
  \sum_{\ww\in\T}
  \frac{Q(\ww)\chi(F(\ww)=C)
   |\im\F|
  }{Q(\T)}
 }
 +1
 \notag
 \\
 &=
 E_{FC}\lrB{
  \lrB{\sum_{\ww\in\T}
   \frac{Q(\ww)\chi(F(\ww)=C)
    |\im\F|
   }{Q(\T)}
  }^2
 }
 -2
 \sum_{\ww\in\T}
 \frac{
  Q(\ww)
  E_{FC}\lrB{\chi(F(\ww)=C)}
  |\im\F|
 }{Q(\T)}
 +1
 \notag
 \\
 &=
 \frac{
  |\im\F|^2
 }{Q(\T)^2}
 E_{FC}\lrB{
  \lrB{\sum_{\ww\in\T}
   Q(\ww)\chi(F(\ww)=C)
  }^2
 }
 -1
 \notag
 \\
 &\leq
 \alpha_F-1
 +
 \frac{
  \lrB{\beta_F+1}
  |\im\F|
  \max_{\ww\in\T}Q(\ww)
 }
 {Q(\T)},
\end{align}
where the inequality comes from (\ref{eq:lemma-multi}).

Finally, the lemma is shown as
\begin{align}
 E_F\lrB{
	\sum_{\cc}
	\left|
	 \frac{Q\lrsb{\T\cap\fC_F(\cc)}}{Q(\T)}
	 -\frac 1{|\im\F|}
	\right|
 }
 &=
 E_{FC}\lrB{
  \left|
   \frac{Q\lrsb{\T\cap\fC_F(C)}
    |\im\F|
   }{Q(\T)}
   -1
  \right|
 }
 \notag
 \\
 &=
 E_{FC}\lrB{
  \sqrt{
   \lrB{
    \frac{
     Q\lrsb{\T\cap\fC_F(C)}
     |\im\F|
    }{Q(\T)}
    -1}^2
  }
 }
 \notag
 \\
 &\leq
 \sqrt{
  E_{FC}\lrB{
   \lrB{\frac{
     Q\lrsb{\T\cap\fC_F(C)}
     |\im\F|
    }{Q(\T)}
    -1}^2
  }
 }
 \notag
 \\
 &\leq
 \sqrt{
  \alpha_F-1
  +
  \frac{
   \lrB{\beta_F+1}
   |\im\F|
   \max_{\ww\in\T}Q(\ww)
  }
  {Q(\T)}
 },
 \label{eq:proof-BCP-multi}
\end{align}
where the first inequality comes from the Jensen inequality.
\end{IEEEproof}

The following lemma is a multiple extension of
the {\em collision-resistant property}.
This lemma implies that
there is an assignment such that every bin contains at most one item.
We use the following notations.
For each $i\in\I$, let $\F_i$ be a set
of functions on $\W_i^n$ and $\cc_i\in\im\F_i$.
Let $\W_{\I'}^n\equiv\Prod_{i\in\I'}\W_i^n$ and 
\begin{align*}
 \alpha_{F_{\I'}}
 &\equiv
 \prod_{i\in\I'}\alpha_{F_i}
 \\
 \beta_{F_{\I'}}
 &\equiv
 \prod_{i\in\I'}\lrB{\beta_{F_i}+1}-1,
\end{align*}
where $\prod_{i\in\emptyset}\theta_i\equiv1$.
It should be noted that
\begin{align*}
 \limn \alpha_{F_{\I'}}
 &=1
 \\
 \limn \beta_{F_{\I'}}
 &=0
\end{align*}
for every $\I'\subset\I$
when $(\aalpha_{F_i},\bbeta_{F_i})$ satisfies
(\ref{eq:alpha})
and (\ref{eq:beta}) for all $s\in\cS$.
For $\T\subset\W_{\I}^n$ and $\ww_{\Ipc}\in\W^n_{\Ipc}$,
let $\T_{\Ipc}$ and $\T_{\I'|\Ipc}(\ww_{\Ipc})$ be defined as
\begin{align*}
 &
 \T_{\Ipc}
 \equiv
 \{\ww_{\Ipc}:
  (\ww_{\I'},\ww_{\Ipc})\in\T
  \ \text{for some}\ \ww_{\I'}\in\W_{\I'}
  \}
 \\
 &
 \T_{\I'|\Ipc}(\ww_{\Ipc})
 \equiv
 \{\ww_{\I'}: (\ww_{\I'},\ww_{\Ipc})\in\T\}.
\end{align*}
\begin{lem}[{\cite[Lemma 7 in the extended version]{HASH-BC}}]
\label{lem:mCRP}
For each $i\in\I$, let $\F_i$ be a set of functions on $\W_i^n$
and $p_{F_i}$ be the probability distribution on $\F_i$,
where $(\F_i,p_{F_i})$ satisfies (\ref{eq:hash}).
We assume that random variables $\{F_i\}_{i\in\I}$ are mutually independent.
Then
\begin{align*}
 p_{F_{\I}}\lrsb{\lrb{
	 f_{\I}:
	 \lrB{\T\setminus\{\ww_{\I}\}}\cap\fC_{f_{\I}}(f_{\I}(\ww_{\I}))
	 \neq
   \emptyset
 }}
 \leq
 \sum_{
	\I'\in2^{\I}\setminus\{\emptyset\}
 }
 \frac{
  \alpha_{F_{\I'}}\lrB{\beta_{F_{\Ipc}}+1}
  \oO_{\I'}
 }
 {
  \prod_{i\in\I'}\lrbar{\im\F_i}
 }
 +\beta_{F_{\I}}
\end{align*}
for all $\T\subset\W_{\I}^n$ and $\ww_{\I}\in\W_{\I}^n$,
where
\begin{equation}
 \oO_{\I'}
 \equiv
 \begin{cases}
	1
  &\text{if}\ \I'=\emptyset,
  \\
  \displaystyle\max_{\ww_{\Ipc}\in\T_{\Ipc}}
  \lrbar{\T_{\I'|\Ipc}\lrsb{\ww_{\Ipc}}}
  &\text{if}\ \emptyset\neq\I'\subsetneq\I,
  \\
  |\T|
  &\text{if}\ \I'=\I.
 \end{cases}
 \label{eq:oO}
\end{equation}
\end{lem}
\begin{IEEEproof}
We show this lemma for the completeness of the paper.
Let
$p_{\ww_i,\ww'_i}
\equiv
p_{F_i}\lrsb{\lrb{
  f_i:
  f_i(\ww_i)=f_i(\ww'_i)
}}$
and let $C_{\I}$ be the random variable
corresponding to the uniform distribution on $\Prod_{i\in\I}\im\F_i$.
In the following, we use the relation
\begin{align}
 \sum_{\substack{
   \ww_i\in\W^n_i:
   \\
   p_{\ww_i,\ww'_i}
   >\frac{\alpha_{F_i}}{|\im\F_i|}
 }}
 p_{\ww_i,\ww'_i}
 &=
 \sum_{\substack{
   \ww_i\in\W^n_i\setminus\{\ww'_i\}:
   \\
   p_{\ww_i,\ww'_i}
   >\frac{\alpha_{F_i}}{|\im\F_i|}
 }}
 p_{\ww_i,\ww'_i}
 +
 p_{\ww'_i,\ww'_i}
 \notag
 \\
 &\leq
 \beta_{F_i}+1
 \label{eq:proof-beta}
\end{align}
for all $\ww'_i\in\W_i^n$, which comes from (\ref{eq:hash})
and the fact that $p_{\ww'_i,\ww'_i}=1$.

First, we have
\begin{align}
 \sum_{\substack{
   \ww_{\I}\in\T:
   \\
   p_{\ww_i,\ww'_i}
   \leq\frac{\alpha_{F_i}}{|\im\F_i|}
   \ \text{for all}\  i\in\I'
   \\
   p_{\ww_i,\ww'_i}
   >\frac{\alpha_{F_i}}{|\im\F_i|}
   \ \text{for all}\ i\in\Ipc
 }}
 \prod_{i\in\I} p_{\ww_i,\ww'_i}
 &=
 \sum_{\substack{
   \ww_{\Ipc}\in\T_{\Ipc}:
   \\
   p_{\ww_i,\ww'_i}
   >\frac{\alpha_{F_i}}{|\im\F_i|}
   \\
   \text{for all}\  i\in\Ipc
 }}
 \lrB{\prod_{i\in\Ipc} p_{\ww_i,\ww'_i}}
 \sum_{\substack{
   \ww_{\I'}\in\T_{\I'|\Ipc}(\ww_{\Ipc}):
   \\
   p_{\ww_i,\ww'_i}
   \leq\frac{\alpha_{F_i}}{|\im\F_i|}
   \\
   \text{for all}\ i\in\I'
 }}
 \prod_{i\in\I'}p_{\ww_i,\ww'_i}
 \notag
 \\
 &\leq
 \lrB{\prod_{i\in\I'}\frac{\alpha_{F_i}}{|\im\F_i|}}
 \sum_{\substack{
   \ww_{\Ipc}\in\T_{\Ipc}:
   \\
   p_{\ww_i,\ww'_i}
   >\frac{\alpha_{F_i}}{|\im\F_i|}
   \\
   \text{for all}\ i\in\Ipc
 }}
 \lrB{
  \prod_{i\in\Ipc} p_{\ww_i,\ww'_i}
 }
 \sum_{\substack{
   \ww_{\I'}\in\T_{\I'|\Ipc}(\ww_{\Ipc}):
   \\
   p_{\ww_i,\ww'_i}
   \leq\frac{\alpha_{F_i}}{|\im\F_i|}
   \\
   \text{for all}\ i\in\I'
 }}
 1
 \notag
 \\
 &\leq
 \oO_{\I'}
 \lrB{\prod_{i\in\I'}\frac{\alpha_{F_i}}{|\im\F_i|}}
 \prod_{i\in\Ipc}
 \lrB{
  \sum_{\substack{
    \ww_i\in\W_i^n:
    \\
    p_{\ww_i,\ww'_i}
    >\frac{\alpha_{F_i}}{|\im\F_i|}
  }}
  p_{\ww_i,\ww'_i}
 }
 \notag
 \\
 &\leq
 \oO_{\I'}
 \lrB{\prod_{i\in\I'}\frac{\alpha_{F_i}}{|\im\F_i|}}
 \prod_{i\in\Ipc}
 \lrB{\beta_{F_i}+1}
 \notag
 \\
 &=
 \frac{
  \alpha_{F_{\I'}}\lrB{\beta_{F_{\Ipc}}+1}
  \oO_{\I'}
 }
 {\prod_{i\in\I'}\lrbar{\im\F_i}}
 \label{eq:lemma-mcrp}
\end{align}
for all $(\ww'_{\I},\I')$ satisfying
$\ww'_{\I}\in\T$ and $\emptyset\neq\I'\subsetneq\I$,
where the second inequality comes from (\ref{eq:oO})
and the third inequality comes from (\ref{eq:proof-beta}).
It should be noted that (\ref{eq:lemma-mcrp})
is valid for cases $\I'=\emptyset$ and $\I'=\I$
by letting $\oO_{\emptyset}\equiv 1$
and $\oO_{\I}\equiv |\T|$, respectively, 
because
\begin{align}
 \sum_{\substack{
   \ww_{\I}\in\T:
   \\
   p_{\ww_i,\ww'_i}
   >\frac{\alpha_{F_i}}{|\im\F_i|}
   \ \text{for all}\ i\in\I
 }}
 \prod_{i\in\I} p_{\ww_i,\ww'_i}
 &\leq
 \prod_{i\in\I}
 \lrB{
  \sum_{\substack{
    \ww_i\in\W_i^n:
    \\
    p_{\ww_i,\ww'_i}
    >\frac{\alpha_{F_i}}{|\im\F_i|}
  }}
  p_{\ww_i,\ww'_i}
 }
 \notag
 \\
 &\leq
 \prod_{i\in\I}
 \lrB{\beta_{F_i}+1}
 \notag
 \\
 &=
 \frac{\alpha_{F_{\emptyset}}\lrB{\beta_{F_{\I}}+1}\oO_{\emptyset}}
 {\prod_{i\in\emptyset}\lrbar{\im\F_i}}
 \label{eq:proof-BCP-multi1}
\end{align}
and
\begin{align}
 \sum_{\substack{
   \ww_{\I}\in\T:
   \\
   p_{\ww_i,\ww'_i}
   \leq\frac{\alpha_{F_i}}{|\im\F_i|}
   \ \text{for all}\ i\in\I
 }}
 \prod_{i\in\I} p_{\ww_i,\ww'_i}
 &\leq
 \frac{\alpha_{F_{\I}}|\T|}
 {\prod_{i\in\I}\lrbar{\im\F_i}}
 \notag
 \\
 &=
 \frac{\alpha_{F_{\I}}\lrB{\beta_{F_{\emptyset}}+1}\oO_{\I}}
 {\prod_{i\in\I}\lrbar{\im\F_i}}.
\end{align}
Then we have
\begin{align}
 p_{F_{\I}}\lrsb{\lrb{
   f_{\I}:
   \lrB{\T\setminus\{\ww_{\I}\}}\cap\fC_{F_{\I}}(f_{\I}(\ww_{\I}))
   \neq
   \emptyset
 }}
 &\leq
 \sum_{\ww'_{\I}\in\T\setminus\{\ww_{\I}\}}
 p_{F_{\I}}\lrsb{\lrb{
   f_{\I}:
   f_{\I}(\ww_{\I})=f_{\I}(\ww'_{\I})
 }}
 \notag
 \\
 &=
 \sum_{\ww'_{\I}\in\T\setminus\{\ww_{\I}\}}
 p_{F_{\I}}\lrsb{\lrb{
   f_{\I}:
   f_i(\ww_i)=f_i(\ww'_i)
   \ \text{for all}\ i\in\I
 }}
 \notag
 \\*
 &=
 \sum_{\ww'_{\I}\in\T}
 \prod_{i\in\I} p_{\ww_i,\ww'_i}
 -
 \prod_{i\in\I} p_{\ww_i,\ww_i}
 \notag
 \\
 &=
 \sum_{\I'\subset\I}
 \sum_{\substack{
	 \ww_{\I}\in\T:
   \\
   p_{\ww_i,\ww'_i}
   \leq\frac{\alpha_{F_i}}{|\im\F_i|}
   \ \text{for all}\ i\in\I'
   \\
   p_{\ww_i,\ww'_i}
   >\frac{\alpha_{F_i}}{|\im\F_i|}
   \ \text{for all}\ i\in\Ipc
 }}
 \prod_{i\in\I} p_{\ww_i,\ww'_i}
 -1
 \notag
 \\
 &\leq
 \sum_{\I'\subset\I}
 \frac{\alpha_{F_{\I'}}\lrB{\beta_{F_{\Ipc}}+1}\oO_{\I'}}
 {\prod_{i\in\I}|\im\F_i|}
 -1
 \notag
 \\
 &=
 \sum_{
	\I'\in2^{\I}\setminus\{\emptyset\}
 }
 \frac{
  \alpha_{F_{\I'}}\lrB{\beta_{F_{\Ipc}}+1}\oO_{\I'}
 }
 {\prod_{i\in\I'}|\im\F_i|}
 +
 \beta_{F_{\I}}
\end{align}
for all $\T\subset\W_{\I'}^n$ and $\ww_{\I'}\in\W_{\I'}^n$,
where the third equality comes from the fact that $p_{\ww_i,\ww_i}=1$,
the second inequality comes from (\ref{eq:lemma-mcrp}),
and the last equality comes from the fact that
$\alpha_{F_{\emptyset}}=1$,
$\beta_{F_{\emptyset^{\complement}}}=\beta_{F_{\I}}$,
$\prod_{i\in\emptyset}|\im\F_i|=1$,
and $\oO_{\emptyset}=1$.
\end{IEEEproof}

\subsection{The First and the Second Terms of (\ref{eq:proof-lossy-error})}
\label{sec:channel-bcp}

Here, we elucidate the first and the second terms
on the righthand side of (\ref{eq:proof-lossy-error}).

Assume that $(\F,p_F)$ has the hash property
((\ref{eq:hash}) in Appendix \ref{sec:hash}),
where we omit the dependence of $(\F,p_F)$ on $n$.
Let $\im\F\equiv\bigcup_{f\in\F}\{f(\ww):\ww\in\W^n\}$.
For brevity, we omit the dependence of $W$ and $X$ on $n\in\NN$
and $i\in\I$ when it appears in the subscript of $\mu$.
We omit the dependence of $\alpha$ and $\beta$ on $n\in\NN$ and $i\in\I$.

We can show the first and the second terms of (\ref{eq:proof-lossy-error})
by using the following lemma.
\begin{lem}[{\cite[Eq. (70)]{CRNG}}]
\label{lem:channel-bcp}
Let $\uT$ be defined as
\begin{equation*}
 \uT
 \equiv
 \lrb{
  (\ww,\xx):
  \frac 1n
   \log_2\frac1{
    \mu_{W|X}(\ww|\xx)
   }
   \geq
   \uH(\WW|\XX)-\e
 }.
\end{equation*} 
Then we have
\begin{align*}
 E_{F_i}\lrB{
  \sum_{\substack{
    \xx\in\X^n,
    \ww\in\W^n,
    \cc\in\im\F
  }}
  \mu_X(\xx)
  \lrbar{
   \mu_{W|X}(
    \fC_F(\cc)
    |\xx
   )
   -
   \frac{
    1
   }{
    |\im\F|
   }
  }
 }
 &\leq
 \sqrt{
  \alpha_F-1
  +
  [\beta_F+1]
  |\im\F|
  2^{-n[\uH(\WW|\XX)-\e]}
 }
 +
 2\mu_{WX}(\uT^{\complement}).
\end{align*}
\end{lem}

\begin{IEEEproof}
Let $\uT(\xx)$
be defined as
\begin{equation*}
 \uT(\xx)
 \equiv
 \lrb{
  \ww:
  (\xx,\ww)
  \in\uT
 }.
\end{equation*}
Then we have
\begin{align}
 &
 E_F\lrB{
  \sum_{
   \xx\in\X^n,
   \cc\in\im\F
	}
	\mu_X(\xx)
	\lrbar{
	 \mu_{W|X}(
    \fC_F(\cc)
    |\xx
   )
   -
   \frac{
    1
   }{
    |\im\F|
   }
  }
 }
 \notag
 \\*
 &\leq
 E_F\lrB{
	\sum_{\substack{
    \xx\in\X^n,
    \cc\in\im\F
	}}
  \mu_X(\xx)
	\lrbar{
   \mu_{W|X}(
    \uT(\xx)
    \cap\fC_F(\cc)
    |\xx
   )
   -
   \frac{
    \mu_{W|X}(
     \uT(\xx)
     |\xx
    )
   }{
    |\im\F|
   }
  }
 }
 +
 E_F\lrB{
  \sum_{\substack{
    \xx\in\X^n,
    \cc\in\im\F
  }}
  \mu_{W|X}(
   \uT(\xx)^{\complement}
   \cap\fC_F(\cc)
   |\xx
  )
  \mu_X(\xx)
 }
 \notag
 \\*
 &\quad
 +
 E_F\lrB{
  \sum_{\substack{
    \xx\in\X^n,
    \cc\in\im\F
  }}
  \frac{
   \mu_{W|X}(
    \uT_{W|X}(\xx)^{\complement}
    |\xx
   )
   \mu_X(\xx)
  }{
    |\im\F|
  }
 }
 \notag
 \\
 &=
 \sum_{
  \xx\in\X^n
 }
 \mu_{W|X}(
  \uT(\xx)
  |\xx
 )
 \mu_X(\xx)
 E_F\lrB{
  \sum_{
   \cc\in\im\F
  }
  \lrbar{
   \frac{
    \mu_{W|X}(
     \uT(\xx)
     \cap\fC_F(\cc)
     |\xx
    )
   }{
    \mu_{W|X}(
     \uT(\xx)
     |\xx
    )
   }
   -
   \frac1
   {|\im\F|}
  }
 }
 +
 2
 \sum_{
  \xx\in\X^n
 }
 \mu_{W|X}(
  \uT(\xx)^{\complement}
  |\xx
 )
 \mu_X(\xx)
 \notag
 \\
 &\leq
 \sum_{
  \xx\in\X^n
 }
 \mu_{W|X}(
  \uT(\xx)
  |\xx
 )
 \mu_X(\xx)
 \sqrt{
  \alpha_F-1
  +
  [\beta_F+1]
  |\im\F|
  \frac{
   2^{-n[\uH(\WW|\XX)-\e]}
  }{
   \mu_{W|X}(
    \uT(\xx)
    |\xx
   )
  }
 }
 +
 2\mu_{WX}(\uT^{\complement}),
 \notag
 \\
 &\leq
 \sum_{
  \xx\in\X^n
 }
 \mu_X(\xx)
 \sqrt{
  \alpha_F-1
  +
  [\beta_F+1]|\im\F|
  2^{-n[\uH(\WW|\XX)-\e]}
 }
 +
 2\mu_{WX}(\uT^{\complement})
 \notag
 \\
 &=
 \sqrt{
  \alpha_F-1
  +
  [\beta_F+1]|\im\F|
  2^{-n[\uH(\WW|\XX)-\e]}
 }
 +
 2\mu_{WX}(\uT^{\complement}),
 \label{eq:proof-crng-bcp-5}
\end{align}
where the second inequality comes from
Lemma~\ref{lem:mBCP} in Appendix \ref{sec:hash}
by letting
$\T\equiv\uT(\xx)$ and $Q\equiv\mu_{W|X}(\cdot|\xx)$.
\end{IEEEproof}

\subsection{From the Third to the Fifth Terms of
 (\ref{eq:proof-lossy-error})}
\label{sec:proof-crp}

Here, we show from the third to the fifth terms
on the righthand side of (\ref{eq:proof-lossy-error}).

Let us assume that
ensembles $(\F_i,p_{F_i})$ and $(\G_i,p_{G_i})$
have the hash property ((\ref{eq:hash}) given in Appendix \ref{sec:hash}) for every $i\in\I$,
where their dependence on $n$ is omitted.
For brevity, the above joint ensemble $(\F_i\times\G_i,p_{(F,G)_i})$
is renamed $(\F_i,p_{F_i})$ in the following.
Furthermore, we omit the dependence of $C$, $W$, $\hW$, and $Y$ on $n$,
when it appears in the subscript of $\mu$.

Let us assume that $(\WW_{\I},\YY,\CC_{\I},\hWW_{\I})$
satisfies the Markov chain
\begin{equation}
 W^n_{\I}\markov (C^{(n)}_{\I},Y^n)\markov\hW^n_{\I}
 \label{eq:markov-decoding}
\end{equation}
and
\begin{equation}
 f_i(W^n_i)=C^{(n)}_i
 \label{eq:source-encoding}
\end{equation}
for all $i\in\I$ and $n\in\NN$.
For a given $\e>0$, let $\oT$ be defined as
\begin{align*}
 \oT
 &\equiv
 \lrb{
  (\ww_{\I},\yy):
  \begin{aligned}
   &
   \frac1n\log_2
   \frac1{\mu_{W_{\I'}|W_{\Ipc}Y}
    (\ww_{\I'}|\ww_{\Ipc},\yy)}
   \leq
   \oH(\WW_{\I'}|\WW_{\Ipc},\YY)+\e
   \\
   &\text{for all}\ \I'\in2^{\I}\setminus\{\emptyset\}
  \end{aligned}
 }.
\end{align*}

First, we show the following lemma.
\begin{lem}[{\cite[Eq. (58)]{CRNG-MULTI}}]
\label{lem:tsdecoding}
Let us assume that $\chww_{\I}(\cc_{\I},\yy)$
outputs one of the elements
in $\oT(\yy)\cap\fC_{f_{\I}}(\cc_{\I})$,
where it outputs an arbitrary element of $\W_{\I}^n$
when $\oT(\yy)\cap\fC_{f_{\I}}(\cc_{\I})=\emptyset$.
Then we have
\begin{align*}
 E_{F_{\I}}\lrB{
  \mu_{W_{\I}Y}\lrsb{
   \lrb{
    (\ww_{\I},\yy): \chww_{\I}(F_{\I}(\ww_{\I}),\yy)\neq\ww_{\I}
   }
  }
 }
 &\leq
 \sum_{
	\I'\in2^{\I}\setminus\{\emptyset\}
 }
 \alpha_{F_{\I'}}\lrB{\beta_{F_{\Ipc}}+1}
 \frac{
  2^{
	 n\lrB{\oH(\WW_{\I'}|\WW_{\Ipc},\YY)+\e}
  }
 }{
  \prod_{i\in\I'}|\im\F_i|
 }
 +\beta_{F_{\I}}
 +\mu_{W_{\I}Y}(\oT^{\complement}).
\end{align*}
\end{lem}
\begin{IEEEproof}
Let
$\oT(\yy)
\equiv
\{\ww_{\I}: (\ww_{\I},\yy)\in\oT\}$ and
Assume that $\ww_{\I}\in\oT(\yy)$
and $\chww_{\I}(f_{\I}(\ww_{\I}),\yy)\neq\ww_{\I}$.
Since $\ww_{\I}\in\fC_{f_{\I}}(f_{\I}(\ww_{\I}))$, we have
$\lrB{\oT(\yy)\setminus\{\ww_{\I}\}}
\cap\fC_{f_{\I}}(f_{\I}(\ww_{\I}))\neq\emptyset$.
We have
\begin{align}
 E_{F_{\I}}\lrB{
	\chi(\chww_{\I}(F_{\I}(\ww_{\I}),\yy)\neq\ww_{\I})
 }
 &\leq
 p_{F_{\I}}\lrsb{\lrb{
	 f_{\I}:
	 \lrB{\oT(\yy)\setminus\{\ww_{\I}\}}
	 \cap\C_{f_{\I}}(f_{\I}(\ww_{\I}))\neq\emptyset
 }}
 \notag
 \\
 &\leq
 \sum_{
	\I'\in2^{\I}\setminus\{\emptyset\}
 }
 \frac{
	\alpha_{F_{\I'}}\lrB{\beta_{F_{\Ipc}}+1}
	\oO_{\I'}
 }{
	\prod_{i\in\I'}\lrbar{\im\F_i}
 }
 +\beta_{F_{\I}}
 \notag
 \\
 &\leq
 \sum_{
	\I'\in2^{\I}\setminus\{\emptyset\}
 }
 \alpha_{F_{\I'}}\lrB{\beta_{F_{\Ipc}}+1}
 \frac{
	2^{
	 n\lrB{\oH(\WW_{\I'}|\WW_{\Ipc},\YY)+\e}
	}
 }{
	\prod_{i\in\I'}\lrbar{\im\F_i}
 }
 +\beta_{F_{\I}},
\end{align}
where
the second inequality comes from Lemma~\ref{lem:mCRP} 
in Appendix \ref{sec:hash}
by letting $\T\equiv\oT(\yy)$
and the third inequality comes from the fact that
\begin{equation*}
 \oO_{\I'}
 \leq
 2^{n\lrB{\oH(\WW_{\I'}|\WW_{\Ipc},\YY)+\e}}.
\end{equation*}
We have
\begin{align}
 &
 E_{F_{\I}}\lrB{
	\mu_{W_{\I}Y}\lrsb{
	 \lrb{
		(\ww_{\I},\yy): \chww_{\I}(F_{\I}(\ww_{\I}),\yy)\neq\ww_{\I}
	 }
	}
 }
 \notag
 \\*
 &=
 E_{F_{\I}}\lrB{
	\sum_{
	 \ww_{\I}\in\W^n_{\I},\yy\in\Y^n
	}
	\mu_{W_{\I}Y}(\ww_{\I},\yy)
	\chi(\chww_{\I}(F_{\I}(\ww_{\I}),\yy)\neq \ww_{\I})
 }
 \notag
 \\*
 &=
 \sum_{(\ww_{\I},\yy)\in\oT}
 \mu_{W_{\I}Y}(\ww_{\I},\yy)
 E_{F_{\I}}\lrB{
	\chi(\chww_{\I}(F_{\I}(\zz_{\I}),\yy)\neq\ww_{\I})
 }
 +
 \sum_{(\ww_{\I},\yy)\in\oT^{\complement}}
 \mu_{W_{\I}Y}(\ww_{\I},\yy)
 E_{F_{\I}}\lrB{
	\chi(\chww_{\I}(F_{\I}(\ww_{\I}),\yy)\neq \ww_{\I})
 }
 \notag
 \\
 &\leq
 \sum_{
	\I'\in2^{\I}\setminus\{\emptyset\}
 }
 \alpha_{F_{\I'}}\lrB{\beta_{F_{\Ipc}}+1}
 \frac{
	2^{
	 n\lrB{\oH(\WW_{\I'}|\WW_{\I\setminus\I'},\YY)+\e}
	}
 }{
	\prod_{i\in\I'}\lrbar{\im\F_i}
 }
 +\beta_{F_{\I}}
 +\mu_{Z_{\I}W}(\oT^{\complement}).
\end{align}
\end{IEEEproof}

Next, we introduce the following lemma on the stochastic decision.
\begin{lem}[{
	\cite[Lemma 20 in the extended version]{CRNG-GAS}
	\cite[Corollary 2]{SDECODING}\cite[Lemma 4]{CRNG-CHANNEL}
}]
\label{lem:sdecoding}
Let $(U,V)$ be a pair consisting of state $U\in\U$
and observation $V\in\V$,
where $\mu_{UV}$ is the joint distribution of $(U,V)$.
We make a stochastic decision with $\mu_{U|V}$
guessing state $U$ from $\hU\in\U$,
that is, the joint distribution of $(U,V,\hU)$ is given as
\begin{equation*}
 \mu_{UV\hU}(u,v,\hu)\equiv\mu_{UV}(u,v)\mu_{U|V}(\hu|v).
\end{equation*}
The decision error probability yielded by this rule
is at most twice the decision error probability
of {\em any} (possible stochastic) decision, 
that is,
\begin{align*}
 \sum_{\substack{
	 u\in\U,v\in\V,\hu\in\U:
	 \\
   \hu\neq u
 }}
 \mu_{UV}(u,v)
 \mu_{U|V}(\hu|v)
 &\leq
 2
 \sum_{\substack{
   u\in\U,v\in\V,\chu\in\U:
   \\
   \chu\neq u
 }}
 \mu_{UV}(u,v)
 \mu_{\chU|V}(\chu|v)
\end{align*}
for arbitrary probability distribution $\mu_{\chU|V}$.
\end{lem}
\begin{IEEEproof}
Here, we show this lemma directly for the completeness of this paper.
We have
\begin{align}
 \sum_{\substack{
	 u\in\U,v\in\V,\hu\in\U:
	 \\
	 \hu\neq u
 }}
 \mu_{UV}(u,v)
 \mu_{U|V}(\hu|v)
 &=
 \sum_{\substack{
	 u\in\U,v\in\V
 }}
 \mu_{UV}(u,v)
 [1-\mu_{U|V}(u|v)]
 \notag
 \\
 &=
 \sum_{\substack{
	 u\in\U,v\in\V
 }}
 [\mu_{U|V}(u|v)-\mu_{U|V}(u|v)^2]
 \mu_V(v)
 \notag
 \\
 &\leq
 \sum_{\substack{
	 u\in\U,v\in\V
 }}
 [\mu_{U|V}(u|v)-\mu_{U|V}(u|v)^2]
 \mu_V(v)
 +\sum_{\substack{
	 u\in\U,v\in\V
 }}
 [\mu_{U|V}(u|v)-\mu_{\chU|V}(u|v)]
 \mu_V(v)
 \notag
 \\*
 &\quad
 +\sum_{\substack{
	 u\in\U,v\in\V
 }}
 [\mu_{U|V}(u|v)-\mu_{\chU|V}(u|v)]^2
 \mu_V(v)
 +\sum_{\substack{
	 u\in\U,v\in\V
 }}
 \mu_{\chU|V}(u|v)[1-\mu_{\chU|V}(u|v)]
 \mu_V(v)
 \notag
 \\
 &=
 \sum_{\substack{
	 u\in\U,v\in\V
 }}
 2\mu_{U|V}(u|v)
 [1-\mu_{\chU|V}(u|v)]
 \mu_V(v)
 \notag
 \\
 &=
 2\sum_{\substack{
	 u\in\U,v\in\V
 }}
 \mu_{UV}(u,v)
 [1-\mu_{\chU|V}(u|v)]
 \notag
 \\
 &=
 2\sum_{\substack{
	 u\in\U,v\in\V,\chu\in\U:
	 \\
	 \chu\neq u
 }}
 \mu_{UV}(u,v)
 \mu_{\chU|V}(\chu|v),
\end{align}
where the inequality comes from the fact that
$\sum_{u\in\U}\mu_{U|V}(u|v)=\sum_{u\in\U}\mu_{\chU|V}(u|v)=1$,
and $\mu_{\chU|V}(u|v)\in[0,1]$ for all $u\in\U$ and $v\in\V$.
\end{IEEEproof}

Finally, we show the following lemma,
which we can derive from the third to the fifth terms
of (\ref{eq:proof-lossy-error}).
\begin{lem}[{\cite[Lemma 2 in the extended version]{ICC}}]
\label{lem:channel-crp}
Assume that $(W^n_{\I},Y^n,C^{(n)}_{\I},\hW^n_{\I})$
satisfy (\ref{eq:markov-decoding}) and (\ref{eq:source-encoding}).
Then the expectation of the probability of the event
$\hWW_{\I}\neq\WW_{\I}$ is evaluated as
\begin{align*}
 &
 E_{F_{\I}}\lrB{
  \sum_{\substack{
    \ww_{\I},\yy,\hww_{\I},\cc_{\I}:\\
    \hww_{\I}\neq\ww_{\I}
  }}
  \mu_{\hW_{\I}|C_{\I}Y}
  (\hww_{\I}|F_{\I}(\ww_{\I}),\yy)
  \mu_{W_{\I}Y}(\ww_{\I},\yy)
 }
 \notag
 \\*
 &\leq
 2
 \sum_{
  \I'\in2^{\I}\setminus\{\emptyset\}
 }
 \alpha_{F_{\I'}}\lrB{\beta_{F_{\Ipc}}+1}
 2^{
  -n\lrB{
   \sum_{i\in\I'}r_i-\oH(\WW_{\I'}|\WW_{\Ipc},\YY)
   -\e
  }
 }
 +2\beta_{F_{\I}}
 +2\mu_{W_{\I}Y}(\oT^{\complement}).
\end{align*}
\end{lem}
\begin{IEEEproof}
For given $f_{\I}$,
the joint distribution of $(W_{\I}^n,Y^n,C^{(n)}_{\I})$ is given as
\begin{align}
 \mu_{W_{\I}C_{\I}Y}(\ww_{\I},\cc_{\I},\yy)
 &=
 \chi(f_{\I}(\ww_{\I})=\cc_{\I})
 \mu_{W_{\I}Y}(\ww_{\I},\yy)
 \label{eq:joint-ZICIU}
\end{align}
from (\ref{eq:source-encoding}).
Then we have
\begin{align}
 \mu_{W_{\I}|C_{\I}Y}(\ww_{\I}|\cc_{\I},\yy)
 &\equiv
 \frac{
  \mu_{W_{\I}C_{\I}Y}(\ww_{\I},\cc_{\I},\yy)
 }{
	\sum_{\ww_{\I}}
  \mu_{W_{\I}C_{\I}Y}(\ww_{\I},\cc_{\I},\yy)
 }
 \notag
 \\
 &=
 \frac{
  \mu_{W_{\I}Y}(\ww_{\I},\yy)\chi(f_{\I}(\ww_{\I})=\cc_{\I})
 }{
  \sum_{\ww_{\I}}
  \mu_{W_{\I}Y}(\ww_{\I},\yy)
  \chi(f_{\I}(\ww_{\I})=\cc_{\I})
 }
 \notag
 \\
 &=
 \frac{
  \mu_{W_{\I}Y}(\ww_{\I},\yy)\chi(f_{\I}(\ww_{\I})=\cc_{\I})
 }{
  \sum_{\ww_{\I}}
  \mu_{W_{\I}Y}(\ww_{\I},\yy)
  \chi(f_{\I}(\ww_{\I})=\cc_{\I})
 }
 \notag
 \\
 &=
 \mu_{\hW_{\I}|C_{\I}Y}(\ww_{\I}|\cc_{\I},\yy),
\end{align}
that is, the constrained-random-number generator
is a stochastic decision with
$\mu_{W_{\I}|C_{\I}Y}$.
By letting
\begin{equation*}
 \mu_{\chW_{\I}|C_{\I}Y}(\chww_{\I}|\cc_{\I},\yy)
 \equiv
 \chi(\chww_{\I}(\cc_{\I},\yy)=\chww_{\I}),
\end{equation*}
we have the fact that
\begin{align}
 &
 \sum_{\substack{
   \ww_{\I},\yy,\hww_{\I}:\\
   \hww_{\I}\neq\ww_{\I}
 }}
 \mu_{\hW_{\I}|C_{\I}Y}
 (\hww_{\I}|f_{\I}(\ww_{\I}),\yy)
 \mu_{W_{\I}Y}(\ww_{\I},\yy)
 \notag
 \\*
 &=
  \sum_{\substack{
    \ww_{\I},\yy,\hww_{\I},\cc_{\I}:\\
    \hww_{\I}\neq\ww_{\I}
  }}
  \mu_{\hW_{\I}|C_{\I}Y}(\hww_{\I}|\cc_{\I},\yy)
  \chi(f_{\I}(\ww_{\I})=\cc_{\I})
  \mu_{W_{\I}Y}(\ww_{\I},\yy)
 \notag
 \\
 &=
 \sum_{\substack{
	 \ww_{\I},\yy,\hww_{\I},\cc_{\I}:\\
   \hww_{\I}\neq\ww_{\I}
 }}
 \mu_{W_{\I}|C_{\I}Y}(\hww_{\I}|\cc_{\I},\yy)
 \mu_{W_{\I}C_{\I}Y}(\ww_{\I},\cc_{\I},\yy)
 \notag
 \\
 &\leq
 2
 \sum_{\substack{
	 \ww_{\I},\yy,\hww_{\I},\cc_{\I}:\\
   \chww_{\I}\neq\ww_{\I}
 }}
 \mu_{\chW_{\I}|C_{\I}Y}(\chww_{\I}|\cc_{\I},\yy)
 \mu_{W_{\I}C_{\I}Y}(\ww_{\I},\cc_{\I},\yy)
 \notag
 \\
 &=
 2
 \sum_{\substack{
	 \ww_{\I},\yy,\hww_{\I},\cc_{\I}:\\
   \chww_{\I}\neq\ww_{\I}
 }}
 \mu_{W_{\I}Y}(\ww_{\I},\yy)
 \chi(f_{\I}(\ww_{\I})=\cc_{\I})
 \chi(\chww_{\I}(\cc_{\I},\yy)=\chww_{\I})
 \notag
 \\
 &=
 2
 \sum_{\substack{
   \ww_{\I},\yy:
   \\
   \chww_{\I}(f_{\I}(\ww_{\I}),\yy)\neq\ww_{\I}
 }}
 \mu_{W_{\I}Y}(\ww_{\I},\yy)
 \notag
 \\
 &=
 2
 \mu_{W_{\I}Y}
 \lrsb{
  \lrb{
   (\ww_{\I},\yy):
   \chww_{\I}(f_{\I}(\ww_{\I}),\yy)\neq\ww_{\I}
  }
 },
\end{align}
where the second and the third equality comes from (\ref{eq:joint-ZICIU})
and the inequality comes from Lemma~\ref{lem:sdecoding}.
This yields the fact that
\begin{align}
 &
 E_{F_{\I}}\lrB{
  \sum_{\substack{
    \ww_{\I},\yy,\hww_{\I}:\\
    \hww_{\I}\neq\ww_{\I}
  }}
  \mu_{\hW_{\I}|C_{\I}Y}
  (\hww_{\I}|f_{\I}(\ww_{\I}),\yy)
  \mu_{W_{\I}Y}(\ww_{\I},\yy)
 }
 \notag
 \\*
 &\leq
 2
 E_{F_{\I}}\lrB{
  \mu_{W_{\I}Y}
  \lrsb{
   \lrb{
    (\ww_{\I},\yy):
    \chww_{\I}(F_{\I}(\ww_{\I}),\yy)\neq\ww_{\I}
   }
  }
 }
 \notag
 \\
 &\leq
 2 \sum_{
	\I'\in2^{\I}\setminus\{\emptyset\}
 }
 \alpha_{F_{\I'}}\lrB{\beta_{F_{\Ipc}}+1}
 2^{
  -n\lrB{
   \sum_{i\in\I'}r_i-\oH(\WW_{\I'}|\WW_{\Ipc},\YY)-\e
  }
 }
 +2 \beta_{F_{\I}}
 +2 \mu_{W_{\I}Y}(\oT^{\complement}),
\end{align}
where we use the relation $r_i=\log_2(|\C_i|)/n=\log_2(|\im\F_i|)/n$
in the last inequality.
\end{IEEEproof}

\subsection{Proof of Lemma}
\label{sec:proof-lemma}

\begin{lem}[{\cite[Lemma 19 in the extended version]{CRNG-GAS}}]
\label{lem:diff-prod}
For any sequence $\{\theta_k\}_{k=1}^K$,
we have
\begin{equation}
 \prod_{k=1}^K \theta_k - 1
 =
 \sum_{k=1}^K
 \lrB{\theta_k - 1}
 \prod_{k'=k+1}^K\theta_{k'},
 \label{eq:diff-prod}
\end{equation}
where $\prod_{k'=K+1}^{K}\theta_{k'}\equiv1$.
For any sequence $\{\theta_k\}_{k=1}^K$ of non-negative numbers,
we have
\begin{equation}
 \lrbar{
  \prod_{k=1}^K \theta_k - 1
 }
 \leq
 \sum_{k=1}^K
 \lrbar{\theta_k - 1}
 \prod_{k'=k+1}^K\theta_{k'}.
 \label{eq:diff-prod-ineq}
\end{equation}
\end{lem}
\begin{IEEEproof}
First, we show (\ref{eq:diff-prod}) by induction.
When $K=1$, (\ref{eq:diff-prod}) is trivial.
Assuming that (\ref{eq:diff-prod}) is satisfied,
we have
\begin{align}
 \prod_{k=1}^{K+1} \theta_k - 1
 &=
 \prod_{k=1}^{K+1} \theta_k
 -
 \theta_{K+1}
 +
 \theta_{K+1}
 - 1
 \notag
 \\
 &=
 \lrB{
  \prod_{k=1}^K \theta_k
  -
  1
 }
 \theta_{K+1}
 +
 \lrB{
  \theta_{K+1}
  - 1
 }
 \notag
 \\
 &=
 \lrB{
  \sum_{k=1}^K
  \lrB{\theta_k - 1}
  \prod_{k'=k+1}^K\theta_{k'}
 }
 \theta_{K+1}
 +
 \lrB{
  \theta_{K+1}
  - 1
 }
 \notag
 \\
 &=
 \sum_{k=1}^{K+1}
 \lrB{\theta_k - 1}
 \prod_{k'=k+1}^{K+1}\theta_{k'},
\end{align}
where the third equality comes from the assumption,
and the last equality comes from the fact that
\[
\lrB{\theta_{K+1}-1}\prod_{k'=K+1+1}^{K+1}\theta_{k'}=\theta_{K+1}-1.
\]
Then (\ref{eq:diff-prod}) is shown by induction.

We have (\ref{eq:diff-prod-ineq}) from (\ref{eq:diff-prod}) as
\begin{align}
 \lrbar{\prod_{k=1}^{K+1} \theta_k - 1}
 &=
 \lrbar{
	\sum_{k=1}^{K+1}
	\lrB{\theta_k - 1}
	\prod_{k'=k+1}^{K+1}\theta_{k'}
 }
 \notag
 \\
 &\leq
 \sum_{k=1}^{K+1}
 \lrbar{
	\lrB{\theta_k - 1}
	\prod_{k'=k+1}^{K+1}\theta_{k'}
 }
 \notag
 \\
 &=
 \sum_{k=1}^{K+1}
 \lrbar{\theta_k - 1}
 \prod_{k'=k+1}^{K+1}\theta_{k'},
\end{align}
where
the inequality comes from the triangle inequality
and the last equality comes from the fact that
$\{\theta_k\}_{k=1}^K$ is the sequence of positive numbers.
\end{IEEEproof}

\end{document}